\documentclass[aip,cha,reprint]{revtex4-1}

\usepackage{graphicx}% Include figure files
\usepackage{dcolumn}% Align table columns on decimal point
\usepackage{bm}% bold math
\usepackage[utf8]{inputenc}
\usepackage[T1]{fontenc}
\usepackage{mathptmx}
\usepackage{etoolbox}
\usepackage{graphicx}
\usepackage{float}
\graphicspath{ {./Figures/} }
\usepackage[dvipsnames]{xcolor}

\makeatletter
\def\@email#1#2{%
 \endgroup
 \patchcmd{\titleblock@produce}
  {\frontmatter@RRAPformat}
  {\frontmatter@RRAPformat{\produce@RRAP{*#1\href{mailto:#2}{#2}}}\frontmatter@RRAPformat}
  {}{}
}%
\makeatother
\begin{document}

\preprint{AIP/123-QED}

\title[Internal Aerodynamics of Supersonic Crossflows with Transverse Liquid Injection]{Internal Aerodynamics of Supersonic Crossflows with Transverse Liquid Injection}

\author{Srinivas M V V}
 \affiliation{Mechanical Engineering Department, Indian Institute of Technology Jodhpur, Jodhpur, Rajasthan, 342030, India}
\author{Arun Kumar Rajagopal}
 \email{arunkr@iitj.ac.in}
\affiliation{Mechanical Engineering Department, Indian Institute of Technology Jodhpur, Jodhpur, Rajasthan, 342030, India}

\author{Lebonah B}
\affiliation{ISRO Propulsion Complex, Mahendragiri, Tamil Nadu, 627 133, India}
\author{Jegesh David M}
\affiliation{ISRO Propulsion Complex, Mahendragiri, Tamil Nadu, 627 133, India}

\date{\today}% It is always \today, today,

\begin{abstract}
This study experimentally investigates the internal aerodynamics of transverse liquid injection in a supersonic crossflow (Mach (\textit{M}) = 2.1) using two configurations: single and tandem (8 \textit{mm} spacing) at three injection mass flow rates. Back-lit imaging revealed classical jet breakup phenomena, including surface wave instabilities with increasing amplitudes along the jet boundary, leading to protrusions, breakup into large liquid clumps, and their disintegration into finer droplets under aerodynamic forces. The single injection exhibited the formation of large liquid clumps further downstream compared to the tandem injection. Schlieren imaging showed that at a low momentum flux ratio (\textit{J} = 0.94), both configurations produced regular reflection (RR) of the bow shock wave from the top wall. Increasing \textit{J} to 1.90 resulted in RR for the single injection, while the tandem injection transitioned to Mach reflection (MR). At \textit{J} = 2.67, both configurations exhibited MR. The earlier RR-to-MR transition in tandem injection is attributed to its higher jet penetration and spanwise spread, which reduce the downstream crossflow passage area, acting as a supersonic diffuser and increasing downstream pressure which is favorable for MR transition. Separation zones were observed at the bottom wall due to bow shock wave-boundary layer interaction, and at the side walls due to the interaction of the Mach stem of the MR structure with the walls. These interactions create complex flow regions dominated by vortex structures, significantly influencing the overall flow dynamics.  
 
\end{abstract}

\maketitle

\section{\label{sec:Intro}Introduction}

Jets in crossflow (JIC) are a subject of active research due to their numerous applications in the aviation industry \cite{srinivas2022numerical, peters2024liquid, nayebi2024computational, maikap2023numerical, andreopoulos1984experimental, murugappan2005control, he2024numerical}. A prominent focus in recent decades has been on fuel injection studies for hypersonic air-breathing engines, particularly in the context of scramjet technology. In scramjet engines, air is ingested into the combustion chamber at supersonic speeds, and sustaining the combustion at such speed poses a major challenge for the realization of such engines. Additionally, the residence time available for air-fuel mixing in scramjet combustor is in the order of 1 millisecond, making efficient and rapid mixing of air and fuel a considerable challenge \cite{lee2015challenges}. Another critical challenge in fuel injection for scramjet engines is the reduced fuel penetration caused by the high momentum of the supersonic crossflow. Transverse fuel injection through struts into the supersonic crossflow has been found to be a promising fuel injection strategy to overcome these challenges \cite{liu2024experimental, qiu2024numerical}. In the past, there have been studies that investigated gaseous as well as liquid jets as fuel in a supersonic crossflow \cite{zhao2022numerical, oamjee2020effects, sekar2024liquid, tang2023numerical, cai2024investigation}. The primary advantage of gaseous fuel is its ability to mix with the crossflow more rapidly than liquid fuel, owing to its superior diffusion properties. However, the major drawback of gaseous fuel lies in its lower penetration, which can hamper combustion efficiency. The key advantage of using liquid fuel is its higher density, which enables deeper penetration of the liquid jet into the crossflow, thereby improving mixing and enhancing combustion efficiency \cite{li2023numerical}. As a result, there is increased interest in injecting the fuel in a liquid state for scramjet combustors. However, it is essential to note that liquid fuel combustion involves multiple processes, including liquid column breakup, primary and secondary atomization into fine droplets, thorough mixing within the combustion chamber, and subsequent fuel evaporation—all of which must be completed within a few milliseconds \cite{zhou2023spray}. These sequential stages of liquid breakup before combustion present a significant challenge for effective fuel injection in the liquid phase.\\

\begin{figure}[htbp]
\includegraphics[width=0.45\textwidth]{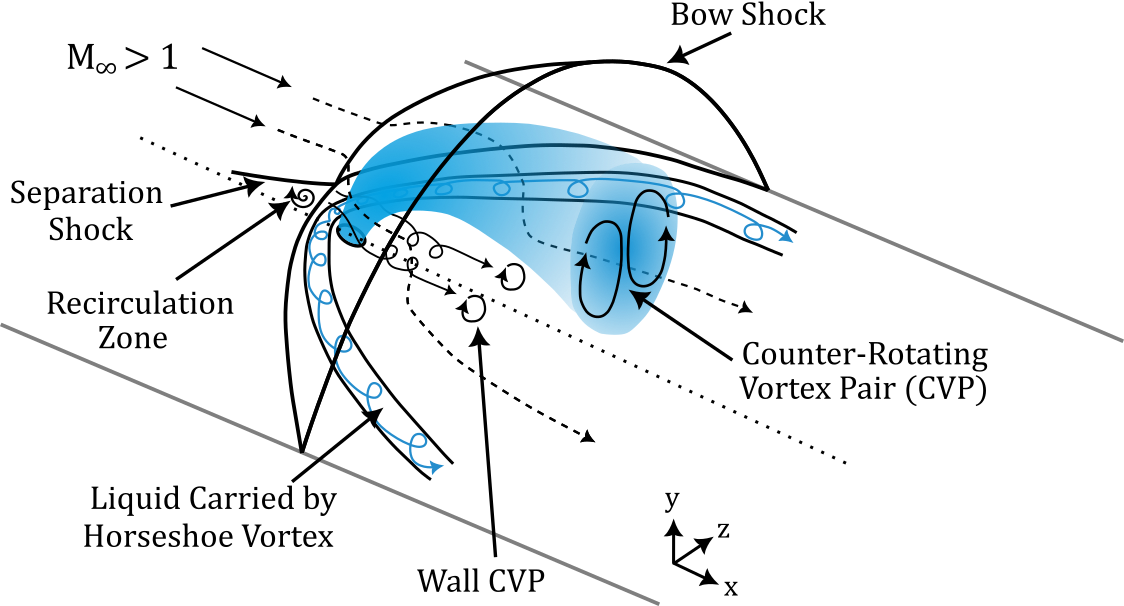}
\caption{\label{fig:bowshock_schematic}A schematic of flow and shock structures produced by transverse liquid injection in a supersonic crossflow}
\end{figure}

\begin{figure*}[htbp]
\includegraphics[width=0.75\textwidth]{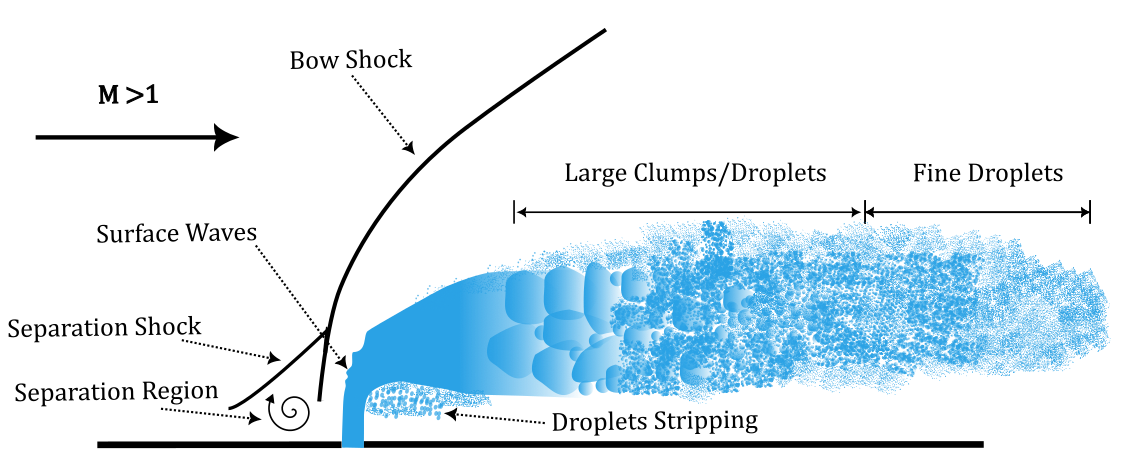}
\caption{\label{fig:schematic}A schematic representation of the liquid jet breakup characteristics in a supersonic crossflow}
\end{figure*}

A schematic of the liquid injection into a supersonic crossflow is shown in Fig. \ref{fig:bowshock_schematic}. The transverse liquid injection in a supersonic crossflow acts as a barrier to the supersonic flow, resulting in the formation of a bow shock wave. The interaction of the bow shock wave with the bottom wall results in boundary layer separation due to adverse pressure gradient across the shock wave \citep{dickmann2009shock}. A separation shock is formed resulting from the development of a separation region as depicted in Fig. \ref{fig:bowshock_schematic}. The recirculation zone formed due to the bow shock wave-boundary layer interaction is carried in the downstream direction by the crossflow, resulting in the formation of a horseshoe vortex. In addition to the horseshoe vortex, two major counter-rotating vortex pairs (CVPs) are formed: (a) the wall CVP, which forms near the bottom wall, and (b) the main CVP, which develops at the core of the liquid spray. The wall CVP arises as the crossflow near the wall region deflects around the liquid jet and moves toward the low-pressure region downstream of the jet. The core spray, which is driven downward due to the transverse pressure gradient interacts with the wall CVP and results in the formation of the main CVP. The formation of the CVPs has been detailed in the numerical study by \citet{Li2019140}. A schematic representation of the liquid jet breakup characteristics in a supersonic crossflow is shown in Fig. \ref{fig:schematic}. The disintegration of the injected liquid begins with the formation of surface waves on the windward side of the liquid surface of the jet. The amplitude of surface waves increases, causing them to protrude into the liquid jet. This process breaks the liquid jet, leading to the formation of larger droplets, which are then further disintegrated into smaller droplets by strong aerodynamic forces \cite{sherman1971breakup, schetz1980wave,li2022spray,srinivas2024fundamental}. 

\citet{less1986transient} carried out several experiments to examine the transient characteristics of liquid jets with normal injection in crossflow with different flow Mach numbers varying from 0.48-3. The frequency of the droplet size fluctuations was found to be in the range of $1-14\ kHz$. Additionally, they noted that the frequency of the liquid jet fracture was associated with the frequency of the surface waves that travel through the column of the liquid jet. \citet{zhao2020structures} experimentally studied the dynamics of the liquid injection in a supersonic crossflow in a confined rectangular duct with an expansion section placed downstream. They identified three distinct stages in the breakup and evolution of the liquid jet: (1) fracture of the liquid column, (2) acceleration of the resulting spray, and (3) spray distribution affected by the compression wave. An experimental study of liquid injection at elevated temperatures by \citet{su2020liquid} demonstrated that both the bow shock and the separation shock significantly influence the dispersion of the spray. Further, the whipping motion of the spray, which contributed to a quicker atomization of the spray, was observed. \citet{wang2014experimental} conducted experimental visualizations of liquid jets, capturing the surface waves near the injection location and vortex structures of the spray in the downstream region. It was also noted that the spray velocity experienced a rapid increase shortly after the injection location, and then the velocity decreased as it moved further downstream. An experimental study to investigate the droplet size distribution and the spray characteristics was performed by \citet{li2021cross}. It has been noticed that both the spray penetration and the spread are linearly dependent on the injector diameter. Additionally, it was observed that the droplet distribution became uniform with the increase in transverse distance. A few research studies\cite{nejad1984effects, theofanous2011aerobreakup} have also focused on investigating the characteristics of liquid jets by varying the properties of the injected liquid.

\begin{figure*}[htbp]
\includegraphics[width=0.9\textwidth]{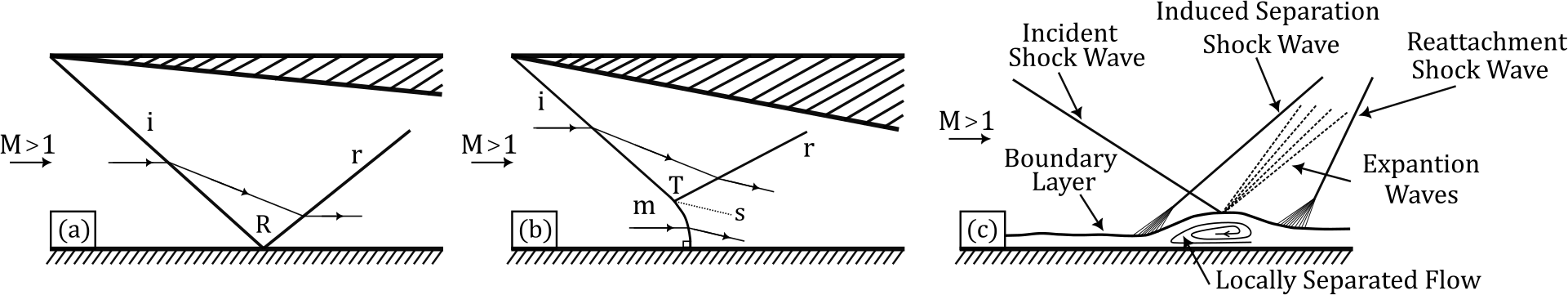}
\caption{\label{fig:rr_mr}Schematic of typical shock wave reflections in a steady supersonic flow (a) Regular Reflection (b) Mach Reflection (c) Shock Wave-Boundary Layer Interaction (SWBLI) ('\textit{i}' - Incident shock wave, '\textit{r}' - Reflected shock wave, '\textit{m}' - Mach stem, '\textit{s}' - Slip line, '\textit{R}' - Point of reflection, '\textit{T}' - Triple point)}
\end{figure*}

In addition to fundamental research on spray characteristics, several studies have concentrated on enhancing mixing and jet penetration. In the past many enhancement techniques were used to improve the mixing in gaseous jet injection in supersonic flow like fluidic oscillators \cite{maikap2024mixing}, tandem injectors \cite{li2017parametric, maikap2024investigation, huang2018mixing}, and shaped injectors\cite{tomioka2011supersonic}. Additionally, limited efforts have been made to enhance the penetration and mixing characteristics of liquid fuel injection. Fuel injection upstream to a cavity leads to better spray mixing inside a cavity that can act as a potential flame-holding site \citep{jiang2024distribution,liu2017dynamics}. \citet{li2017characterization} performed an experimental investigation of kerosene injection with a cavity and it has been observed that the entrainment of the liquid fuel is strongly influenced by the cavity shear layer. Further, it was observed that an increase in the injection pressure resulted in reduced fuel entrainment into the cavity. \citet{hu2019effects} performed both experimental and numerical investigations to study the effect of liquid injection followed by a gaseous jet in a supersonic crossflow. Under identical pressure drop conditions for both gaseous and liquid jets, it was observed that increasing the distance between the gaseous and liquid jets led to a reduction in the penetration of the liquid jet. \citet{bushnell1970experimental} investigated tandem liquid jets injected in a hypersonic crossflow. The results indicated that the jet penetration obtained by the tandem injectors positioned in the stream-wise direction was higher compared to a single injector. \citet{sathiyamoorthy2020penetration} conducted a comparison of the penetration of liquid jets for single and tandem injectors using the shadowgraph imaging technique. Similar to the study by \citet{bushnell1970experimental}, they observed that the tandem injectors exhibited greater jet penetration compared to the single injector. Furthermore, the parametric analysis indicated that increasing the distance between the jets resulted in greater penetration. Apart from the above-discussed enhancement techniques, few studies have explored additional enhancement methods such as pulsating liquid jets \cite{zhu2019primary}, fuel injection behind a pylon \cite{verma2020liquid}, and strut-assisted fuel injection \cite{vinogradov1993injection}.

In addition to the jet breakup characteristics, the injection of fuel in a supersonic crossflow generates complex shock structures and different vortex patterns \cite{rasheed2020numerical, santiago1997crossflow}. The interaction of the bow shock with the walls of a confined duct can cause multiple shock reflections, significantly affecting liquid penetration and mixing, as well as leading to a substantial total pressure loss. Classical literature reported that shock reflections can be classified into two categories namely: Regular Reflection (RR) and Mach Reflection (MR) \cite{ben2007shock}. A schematic illustration of RR and MR in an inviscid flow is shown in Fig. \ref{fig:rr_mr}(a) and Fig. \ref{fig:rr_mr}(b), respectively. A typical, RR structure involves two-shock discontinuity: an incident shock wave, referred to as '\textit{i}', and a reflected shock wave, referred to as '\textit{r}'. The incident and reflected shock waves converge at a point known as the reflection point, denoted by '\textit{R}' as shown in Fig. \ref{fig:rr_mr}(a). The schematic of MR is shown in Fig. \ref{fig:rr_mr}(b). Mach reflection involves the interaction of three distinct shock waves: the incident shock wave (\textit{i}), the reflected shock wave (\textit{r}), and the Mach stem (\textit{m}), and a flow discontinuity, slipstream denoted by '\textit{s}' all converging at a point known as the triple point, denoted by '\textit{T}'. The Mach stem stands perpendicular very close to the wall surface and behaves similarly to a normal shock wave, thereby causing the flow to be subsonic as it passes the Mach stem. The consideration of the viscous effects can modify the shock reflection structures due to boundary layer thickening and subsequent flow separation. A typical representation of the regular reflection structure produced in a viscous flow in shown in Fig. \ref{fig:rr_mr}(c). The interaction of the incident shock wave with the boundary layer, results in the thickening of the boundary layer which can lead to the formation of a local separation region. This thickening of the boundary layer generates compression waves near the separation region, which coalesce to form an induced separation shock wave away from the surface. Above the separation region, expansion waves are developed, while in the downstream region, additional compression waves converge to form a reattachment shock wave. Numerous studies have concentrated on the complex physics of shock reflections that take place in supersonic flows within confined ducts \cite{kumar2017shock, azevedo1989analytic}. A few studies have investigated the effect of shock interaction on the liquid jet characteristics \cite{medipati2023liquid, zhang2025study, sebastian2024transition}. \citet{medipati2023liquid} investigated the unsteadiness during liquid injection and observed that the frequency of jet oscillations closely aligned with the typical frequencies found in SWBLI, indicating that the unsteadiness in the liquid jet can be attributed to SWBLI. \citet{zhang2025study} observed that the penetration of the liquid jet and the spray distribution in the spanwise direction is enhanced due to the interaction of the shock wave with the liquid jet. \citet{sebastian2024transition} experimentally investigated the unsteady nature of the shock-shock interactions between an oblique shock produced by a shock generator and the bow shock wave formed due to the liquid injection. It has been reported that various types of shock reflections as classified by \citet{edney1968anomalous} can be seen depending on the momentum flux ratios.\\

It can be seen that past studies have provided considerable insight into the fundamental characteristics of liquid jet injection into supersonic crossflows. Additionally, several jet penetration and mixing enhancement techniques have also been investigated. However, to the best of the authors' knowledge, the shock reflection characteristics and its dynamics during the liquid injection have not been fully explored, particularly with tandem injection configurations. Further, the shock wave reflections due to the injection of liquid jet become more prominent with a low aspect ratio combustors, significantly influencing the spray mixing and penetration characteristics. Moreover, shock wave-boundary layer interactions can induce flow separation and vortex formation, significantly complicating the internal aerodynamics, and a comprehensive investigation into these separation characteristics remains largely unexplored. Therefore, the current study focuses on investigating the nature of shock reflections and their transitions arising due to the liquid injection in a confined rectangular duct with a single injector and tandem injector. Additionally, the liquid break-up characteristics with tandem injection, the jet penetration and spread with tandem injection and its influence on the shock structures, and the complex flow separation regions formed due to the shock wave boundary layer interactions are also experimentally investigated.

\begin{figure*}[htbp]
\includegraphics[width=0.9\textwidth]{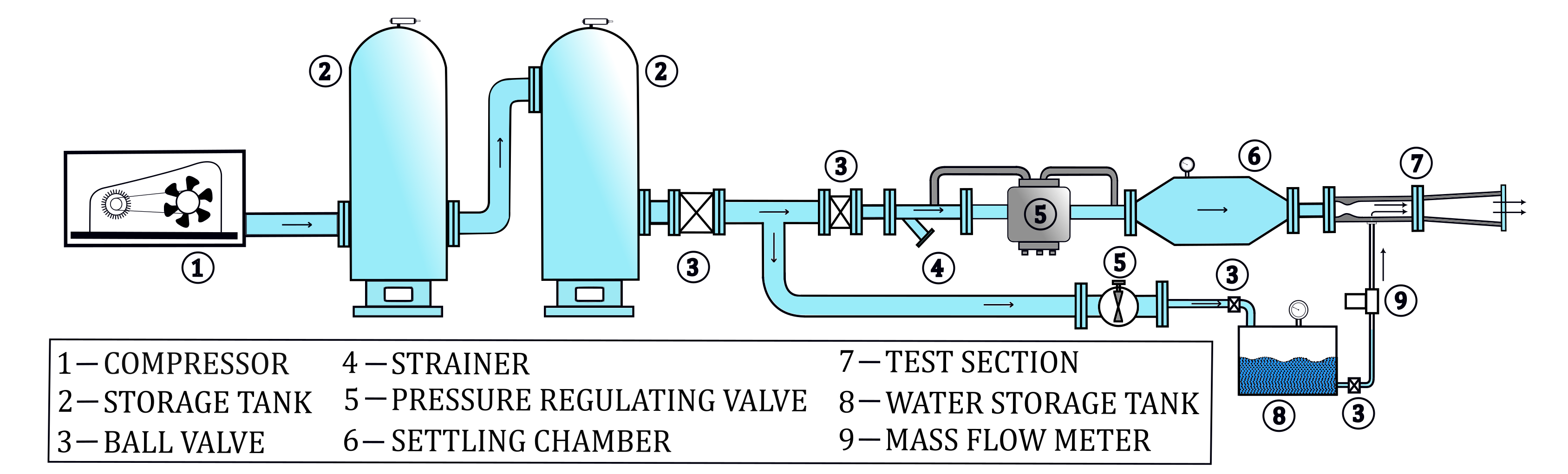}
\caption{\label{fig:setup}Schematic of the liquid injection experimental setup}
\end{figure*}

\begin{figure*}[htbp]
\includegraphics[width=0.85\textwidth]{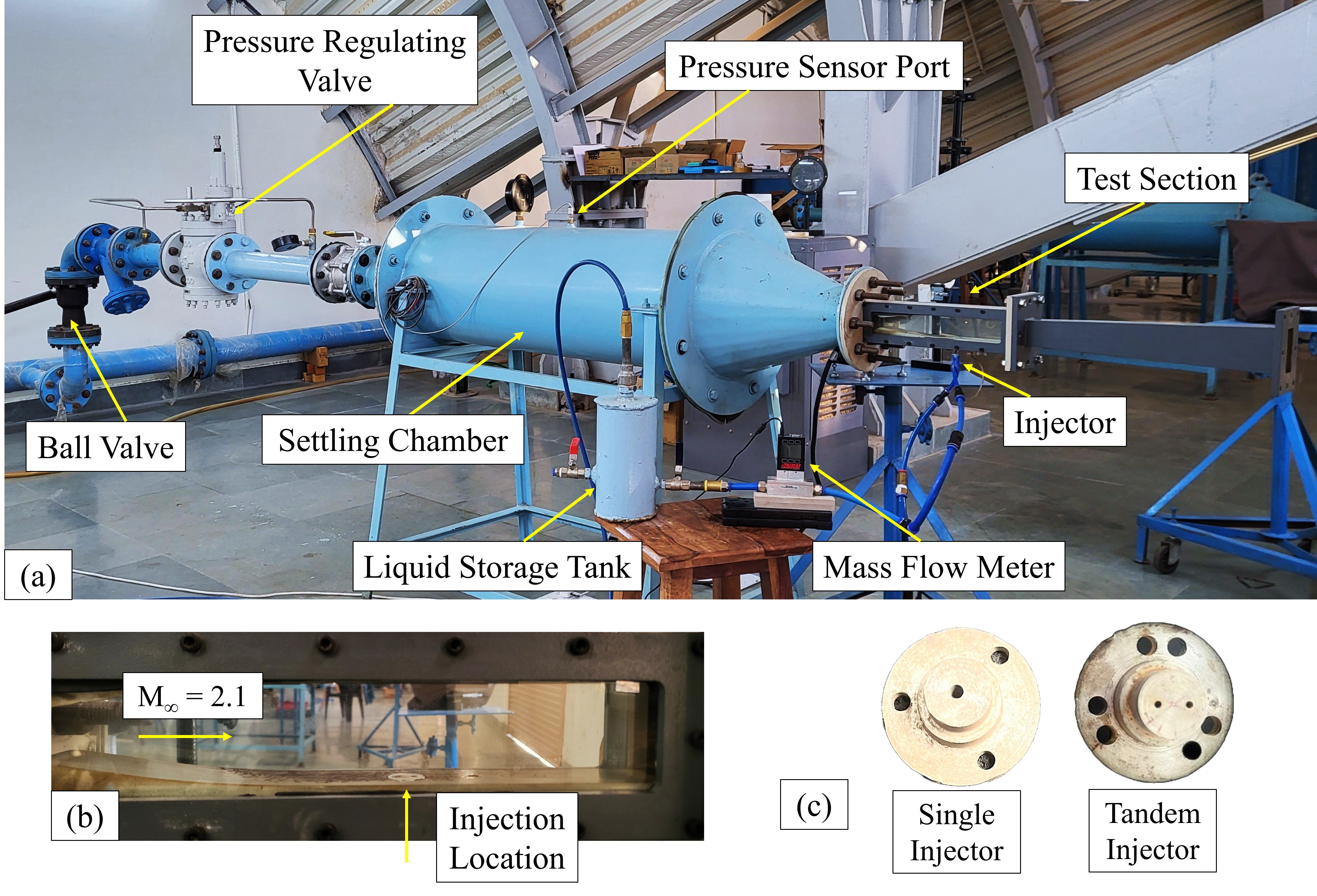}
\caption{\label{fig:exp_setup}(a) Supersonic wind tunnel with pressure feed liquid injection system (b) Wind tunnel test section used for the liquid injection experiments (c) Single and Tandem injectors employed in the study}
\end{figure*}

\section{\label{sec:exp_meth}Experimental Methodology}
\subsection{\label{sec:tunnel_details}Supersonic Wind Tunnel Details}

The experiments for the present study were carried out using the supersonic blowdown wind tunnel facility available at the Shock Waves and High-Speed flows (SWAHS) Lab, IIT Jodhpur, India and the schematic of the facility is shown in Fig. \ref{fig:setup}. High-pressure storage facility with 30 \textit{bar} pressure and $6\ m\textsuperscript{3}$ storage capacity supplies compressed air to the settling chamber of the wind tunnel. A pressure regulating valve (PRV) is used to maintain a constant stagnation pressure (3.4 \textit{bar}) in the settling chamber. A Keller-based pressure sensor (PAA-21Y) with an accuracy of ±0.25\% FS and a limiting frequency of 2 \textit{kHz} has been used to measure the stagnation pressure inside the settling chamber. The wind tunnel characterization experiments were performed to calibrate the wind tunnel and the details of the characterization experiments were reported by \citet{maikap2024mixing}. The actual experimental setup is shown in Fig. \ref{fig:exp_setup}(a). The wind tunnel features a convergent-divergent (C-D) nozzle designed using the method of characteristics (MOC) capable of generating a steady supersonic flow at a designed Mach number of 2.2, with throat (\textit{t}) and exit heights of $25\ mm$ and $50\ mm$, respectively. The C-D nozzle is followed by a constant area test section of length (\textit{L}) $300\ mm$ with a span (\textit{w}) and height (\textit{h}) of $40\ mm$ and $50\ mm$ respectively and a hydraulic diameter (\textit{d\textsubscript{h,c}}) of 44.44 \textit{mm}. The span (\textit{w}) and the height (\textit{h}) of the current test section are non-dimensionalized with the throat height (\textit{t}), resulting in non-dimensional geometric parameters of $w/t = 1.6$ and $h/t = 2$, respectively. The stagnation pressure and the static pressure in the test section are measured to compute the tunnel Mach number and the details of the Mach number measurement are given in the appendix \ref{sec:Mach_measure}. The error linked with the Mach number measurement is determined through multiple repetitions of the experiments, followed by the computation of the standard deviation of the measured Mach number values. The measured Mach number is $2.1 \pm 0.04$ and the deviation from the theoretical Mach number of 2.2 can be attributed to the boundary layer development, which was not considered during the calculation of the C-D nozzle dimensions using MOC. A diverging section is connected at the end of the test section for the pressure recovery. The details of the supersonic crossflow parameters used in the current study are shown in Table \ref{tab:cross_details}. 

The total temperature (T\textsubscript{o}) is considered to be 300 \textit{K} and the static temperature has been calculated using the isentropic flow relations for \textit{M} = 2.1. The density of the supersonic crossflow has been calculated using Eq. \ref{eq:one}, where \textit{R} is the gas constant.  
\begin{equation}
\rho_c = \frac{P\textsubscript{s}}{R\times T\textsubscript{s}}
\label{eq:one}
\end{equation}
The velocity of the supersonic crossflow has been calculated using Eq. \ref{eq:two}, where the ratio of specific heats of air ($\gamma$) is considered to be 1.4.
\begin{equation}
U_c = M \times \sqrt{\gamma \times R \times T_s}
\label{eq:two}
\end{equation}

\begin{table}
\caption{\label{tab:cross_details}Details of the supersonic crossflow parameters}
\begin{ruledtabular}
\begin{tabular}{ccccccc}
 \textit{M}& $ P\textsubscript{o,1}\ (bar)$ &$ P\textsubscript{s}\ (bar) $ \footnotemark[1]&$ T\textsubscript{o,1}\ (K)$& $ T\textsubscript{s}\ (K) $ \footnotemark[2] &$\rho_c\ (kg/m\textsuperscript{3})$&$ U_c\ (m/s)$\\
\hline
2.1& 3.4 & 0.38 & 300 & 159.4 & 0.83 & 531.46 \\
\end{tabular}
\end{ruledtabular}
\footnotetext[1]{$ P\textsubscript{s} $ has been measured experimentally}
\footnotetext[2]{$ T\textsubscript{s}$ has been calculated from $ T\textsubscript{s}/T\textsubscript{o,1}$ using isentropic flow relations for $M = 2.1$}
\end{table}

\subsection{\label{sec:testsection}Details of Injection Configurations}
\begin{figure}[htbp]
\includegraphics[width=0.47\textwidth]{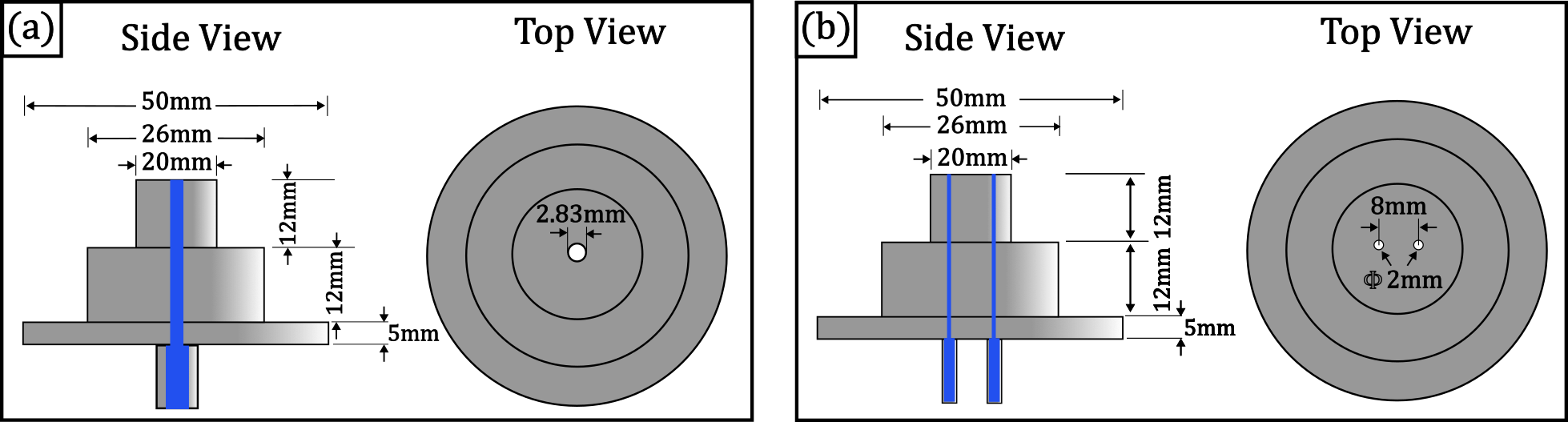}
\caption{\label{fig:inj_comp}Schematic of the side view and top view with dimensions of the injectors used in the current study. (a) Single injector (b) Tandem injector with 8 \textit{mm} spacing}
\end{figure}

For this study, two different injectors were employed: (a) an injector with a single circular port, and (b) an injector with two circular ports oriented in a stream-wise tandem fashion with a spacing of $8\ mm$. For the tandem injector, a port diameter ($d\textsubscript{tandem}$) of $2\ mm$ was used for each port, whereas for the single injector, the port diameter ($d\textsubscript{single}$) used was $2.83\ mm$, such that the total cross-sectional area of the injector ports remains the same for both single and tandem injection configurations ($(\pi/4)\times d\textsuperscript{2}\textsubscript{single} = 2 \times(\pi/4)\times d\textsuperscript{2}\textsubscript{tandem}$). This will ensure that both the injectors will deliver the same mass flow rate of the liquid under identical injection pressure conditions. The injection ports are drilled in a circular insert and are flush mounted at the test section bottom wall region at a distance of 40 \textit{mm} from the nozzle exit as shown in Fig. \ref{fig:exp_setup}(b) and \ref{fig:exp_setup}(c). A schematic of the injectors are also illustrated in Fig. \ref{fig:inj_comp}. The injected liquid is pressure-fed from a liquid storage tank using compressed air from the main air supply line. The injection mass flow rate is controlled by varying the pressure within the liquid storage tank using a separate pressure regulating valve. The mass flow rate of the injected liquid is measured using an Alicat make mass flow meter with a range of 0-10 LPM. The details of the injection supply line can be seen in Fig. \ref{fig:setup} and Fig. \ref{fig:exp_setup}(a). The velocity of the liquid jet ($U_l$) at the exit of the injection orifice was calculated using the mass flow rate obtained from the mass flow meter, using Eq. \ref{eq:three}. For the single injection, the velocity is calculated based on the total injected mass flow rate and an injector diameter of $d_{injector} = 2.83\ mm$. However, for the tandem injection configuration, the exit velocity of each liquid jet was calculated using half of the total injected mass flow rate, as the liquid exits through two separate orifices, each with a diameter of $d_{injector} = 2\ mm$. 
\begin{equation}
U_l = \frac{\dot{m}_l}{\rho_l \times (\frac{\pi}{4}d_{injector}^{2})} 
\label{eq:three}
\end{equation}
The Reynolds number of the liquid jet at the exit of the orifice is calculated using Eq. \ref{eq:four}, where $\mu$ is the dynamic viscosity of liquid.
\begin{equation}
Re_l = \frac{\rho_l \times \ U_l \times d_{injector}}{\mu}
\label{eq:four}
\end{equation}
The Weber number based on the liquid is calculated using Eq. \ref{eq:five}, where $\sigma$ is the surface tension of the liquid. 
\begin{equation}
We_l = \frac{\rho_l \times U_l^2 \times d_{injector}}{\sigma}
\label{eq:five}
\end{equation}
The value of the momentum flux for different cases has been computed using Eq.~(\ref{eq:six})
\begin{equation}
J = \frac{\rho_l U^{2}_l}{\rho_{c} U^2_{c}}
\label{eq:six}
\end{equation}
The liquid jet parameters for different test cases have been shown in Table \ref{tab:details_inj}.

\begin{table*}
\caption{\label{tab:details_inj}Details of the different cases considered in the present study}
\begin{ruledtabular}
\begin{tabular}{ccccccccc}
Case Number & Injector & P\textsubscript{inj}\footnote{Absolute pressure of injection} $(bar)$ & $d\textsubscript{\textit{injector}}$ $(mm)$& $\dot{m}\textsubscript{\textit{l}}$\footnotemark[2] $(kg/s)$ & $U\textsubscript{\textit{l}}$\footnotemark[2] $(m/s)$ & $Re\textsubscript{\textit{l}}$\footnotemark[2] & $We\textsubscript{l}$\footnotemark[2] & $J$ \\ \hline
 C\textsubscript{1} & Single & 2 & 2.83 & 0.08667 & 13.77 & 47030.02 & 6976.19 & 0.94\\
C\textsubscript{2} & Single & 3 & 2.83 & 0.12333 & 19.60 & 65842.02 & 13673.34 & 1.90\\
C\textsubscript{3} & Single & 5 & 2.83 & 0.14633 & 23.26 & 75248.03 & 17859.05 & 2.67\\
 C\textsubscript{4} & Tandem & 2 & 2 & 0.04333\footnotemark[3] & 13.25 & 33273.73 & 4941.15 & 0.94\\
C\textsubscript{5} & Tandem & 3 & 2 & 0.06167\footnotemark[3] & 18.54 & 46583.23 & 9684.65 & 1.90\\
C\textsubscript{6} & Tandem & 5 & 2 & 0.07316\footnotemark[3] & 21.19 & 53237.98 & 12649.34 & 2.67\\
\footnotetext[2]{\textsubscript{\textit{l}} - liquid jet parameter}
\footnotetext[3]{Considering that the $\dot{m}\textsubscript{\textit{l}}$ through each of the injector is half of the total $\dot{m}\textsubscript{\textit{l}}$ injected}
\end{tabular}
\end{ruledtabular}
\end{table*}

\subsection{\label{sec:optic}Details of Experimental Measurement Techniques}
This section provides an overview of the visualization techniques employed in the present work, including back-lit imaging, the Schlieren technique, and the pressure measurement for the investigation of different flow characteristics.

\subsubsection{\label{sec:back-lit}Back-lit Imaging and Schlieren Visualization}
Back-lit imaging technique was employed in the current study to visualize the dynamics of the injected liquid jet. The experiments for the back-lit imaging were performed using a Newport light source with a quartz tungsten halogen filament set at 95 \textit{W}. For the back-lit imaging, optically accessible windows are provided on all four sides of the test section. Glass plates of 12 \textit{mm} thickness were used for the side walls which enables the visualization from the sides. For the top-view back-lit imaging, rectangular slots were machined into the top and bottom walls, and acrylic plates were fitted into these slots. Schlieren imaging technique has been used to visualize the density gradients present in the test section. The same light source used for the back-lit imaging has been used for the Schlieren visualization. The light emitted by the light source is directed towards a rectangular slit to create an extended light source. This rectangular slit is placed at the focus of the first concave mirror having a diameter of 200 \textit{mm} and a focal length of 1000 \textit{mm} to generate a parallel beam of light which is directed onto the test section. Further, the light beam passes through the test section falling on the second mirror having the same diameter and focal length as the first mirror. After reflecting off the second concave mirror, the light beam is bisected by a knife edge positioned at the mirror's focal point. The Schlieren images have been captured using a Chronos-made 2.1 HD high-speed camera. A ZEISS Milvus macro lens having a focal length of 100 \textit{mm} and an aperture range of $f/2.0\ -f/22$ has been used. The far-view images were captured with a resolution of 1280 x 720 and 2142 frames per second (FPS) and the close-up images of the liquid jet have been captured with an image resolution of 320 x 240 and 16682 FPS. The exposure time of the camera sensor was set to be 15 $\mu$s.

\subsubsection{\label{sec:pressure}Pressure Measurement}
The static pressure within the test section is measured using circular pressure ports of 1 \textit{mm} diameter at different stream-wise locations within the test section. Two different models of Keller-based piezoresistive pressure sensors were used to measure the static pressure inside the test section. The Keller model PAA-M5 HB is a voltage output pressure sensor with a range of 0-10 \textit{bar}. It has an accuracy of ±0.1\% FS and an error band of ±0.5\% FS, along with a limiting frequency of 50 \textit{kHz}. Another Keller-based pressure sensor PA-21 PHP model is also a voltage output pressure sensor with a range of 0-10 \textit{bar}. It has an accuracy of $\leq \pm 0.5 \% $ FS and a total error band of $\leq \pm 1 \% $ FS with a limiting frequency of 2 \textit{kHz}. Three different pressure ports: P\textsubscript{1}, P\textsubscript{2}, and P\textsubscript{3} have been chosen to measure the pressure within the test section located at 60 \textit{mm}, 110 \textit{mm}, and 118 \textit{mm} from the injection location, respectively. The voltage output from the pressure sensors was captured using the NI data acquisition system with a sampling frequency of $2\ kHz$, and the total number of samples collected was 20,000. This ensured that the data samples were collected throughout the run time of the experiment.

\section{\label{sec:results}Results and Discussion}
\subsection{\label{sec:jet_char}Liquid Jet Characteristics}

\begin{figure}[!htbp]
\includegraphics[width=0.47\textwidth]{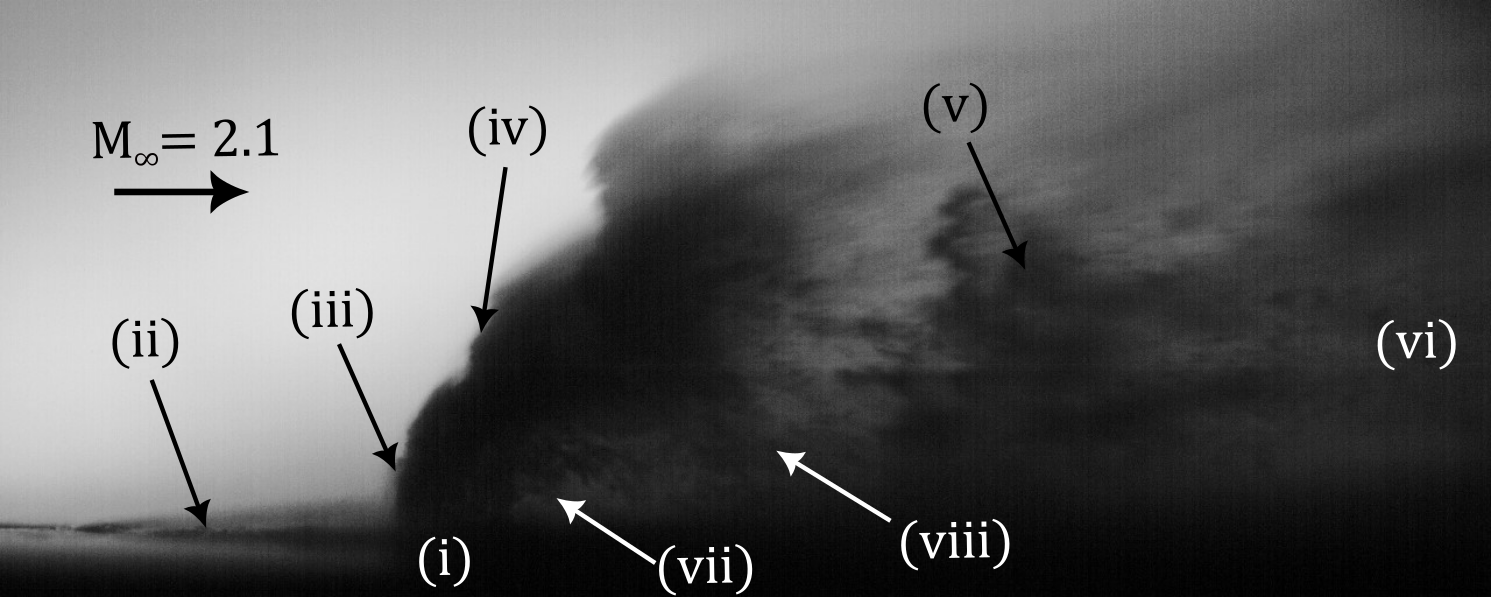}
  \caption{\label{fig:liq_jet}Back-lit image captured at a particular time instant for single injection configuration at $J = 0.94$ showing various zones of the liquid jet (i) Liquid jet (ii) Liquid entrainment (iii) Surface Waves on the liquid jet (iv) Protrusion on the liquid jet boundary (v) Formation of large liquid clumps (vi) Fine droplets (vii) stripping of droplets from the leeward side of the jet (viii) Liquid trailing towards the bottom wall}
\end{figure}

This section examines the breakup characteristics of the pressurized liquid jet injected into a Mach 2.1 supersonic crossflow. Back-lit imaging experiments were conducted for both single and tandem injection to compare their jet breakup behaviors. Fig. \ref{fig:liq_jet} presents an instantaneous closeup view of a single liquid jet injected with a momentum flux ratio, $J = 0.94$, captured using back-lit imaging. Various characteristics of the liquid jet were observed and categorized into distinct zones and are shown in Fig. \ref{fig:liq_jet}. Very close to the injector, the injected liquid jet behaves as a liquid column and is marked as zone (i). Zone (ii) represents the liquid entrainment region upstream of the injection location, caused by the adverse pressure gradient across the bow shock wave interacting with the boundary layer and forming a separation region. As outlined in the introduction section \ref{sec:Intro}, the injection of a liquid jet into a supersonic crossflow, generates a bow shock wave. The interaction of this bow shock wave with the boundary layer leads to the formation of a separation region upstream of the bow shock. The adverse pressure gradient created across the bow shock wave causes the liquid to entrain into the separation region. Zone (iii) shows the formation of the surface wave instabilities on the windward side of the liquid jet. These surface waves are formed as a result of Kelvin-Helmholtz (K-H) instability and Rayleigh-Taylor (R-T) instability \cite{yaozhi2023review}. The amplitude of the surface waves increases as they propagate along the jet boundary, eventually giving rise to a protrusion structure along the surface of the jet which is shown as zone (iv). This protrusion structure initiates the breakup of the liquid jet and results in the formation of large liquid clumps depicted by zone (v) in Fig. \ref{fig:liq_jet}. The liquid clumps are carried downstream by the crossflow, where the shearing action produced by the aerodynamic forces breaks the larger clumps into smaller droplets that mix with the supersonic crossflow, forming a zone (vi). The shearing action also occurs around the liquid jet very close to the bottom wall resulting in the stripping of smaller droplets from the leeward side of the liquid jet, which is marked as zone (vii). A low-pressure region forms in the wake of the liquid jet, and as the crossflow deflects around the jet, it generates a small counter-rotating vortex pair near the wall. This vortex pair effectively drags the denser droplets located in the lower region of the liquid jet towards the bottom wall and this process is termed as liquid trailing phenomenon \citep{li2017numerical}. The liquid trailing zone has been depicted by zone (viii) in Fig. \ref{fig:liq_jet}.

\begin{figure}[!htbp]
\includegraphics[width=0.47\textwidth]{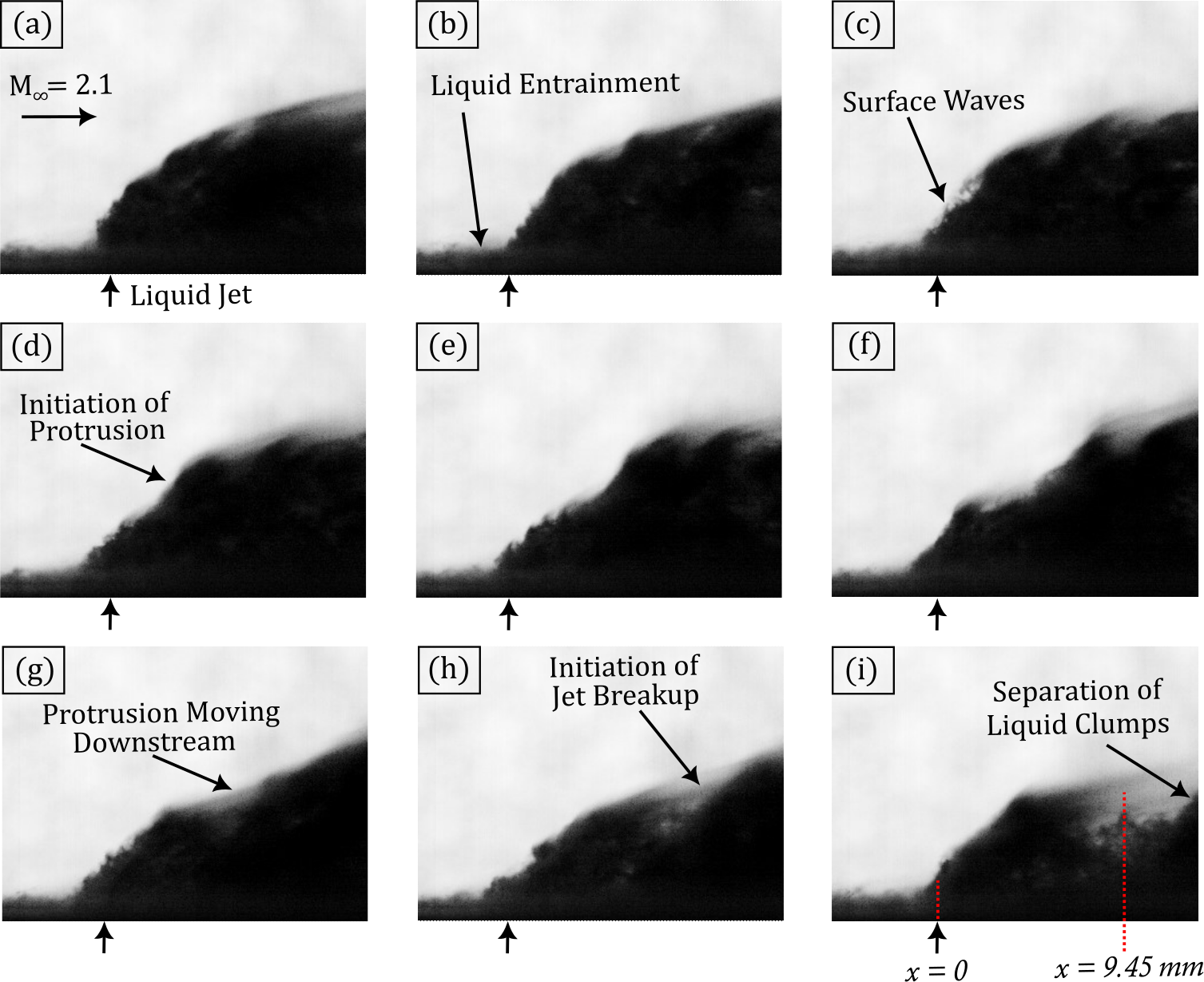}
  \caption{\label{fig:sequential_single}Sequential back-lit images captured with the single liquid injection at $J = 0.94$. (a) $t\textsubscript{0} = 0\ \mu s$ (b) $t\textsubscript{1} = 60\ \mu s$ (c) $t\textsubscript{2} = 120\ \mu s$ (d) $t\textsubscript{3} = 180\ \mu s$ (e) $t\textsubscript{4} = 240\ \mu s$ (f) $t\textsubscript{5} = 300\ \mu s$ (g) $t\textsubscript{6} = 360\ \mu s$ (h) $t\textsubscript{7} = 420\ \mu s$ (i) $t\textsubscript{8} = 480\ \mu s$}
\end{figure}

Fig. \ref{fig:sequential_single} presents a close-up view of the transient evolution of the liquid jet for the single injection case at $J = 0.94$. Several crucial features pertaining to the liquid jet like the liquid entrainment (Fig. \ref{fig:sequential_single}(a) and  \ref{fig:sequential_single}(b)), surface waves (Fig. \ref{fig:sequential_single}(c)), and initiation of the breakup (Fig. \ref{fig:sequential_single}(h)) have been observed with the close-up view images.  The amplitude of the surface waves increases and eventually evolves into a protrusion structure, that initiates the liquid breakup as seen in Fig. \ref{fig:sequential_single}(d). The protrusion structure extends deeply into the liquid jet while simultaneously propagating downstream along the jet boundary as observed in Fig. \ref{fig:sequential_single}(e), \ref{fig:sequential_single}(f), and \ref{fig:sequential_single}(g). The protrusion structure ultimately triggers the breakup of the liquid jet after traveling along the liquid jet boundary, as depicted in Fig. \ref{fig:sequential_single}(h). Following the jet breakup, large liquid clumps separate from the main jet, as shown in Fig. \ref{fig:sequential_single}(i). The breakup of the liquid jet occurs at around $x = 9.45\ mm$ measured from the start of the injection location for the single injection case.

\begin{figure}[!htbp]
\includegraphics[width=0.47\textwidth]{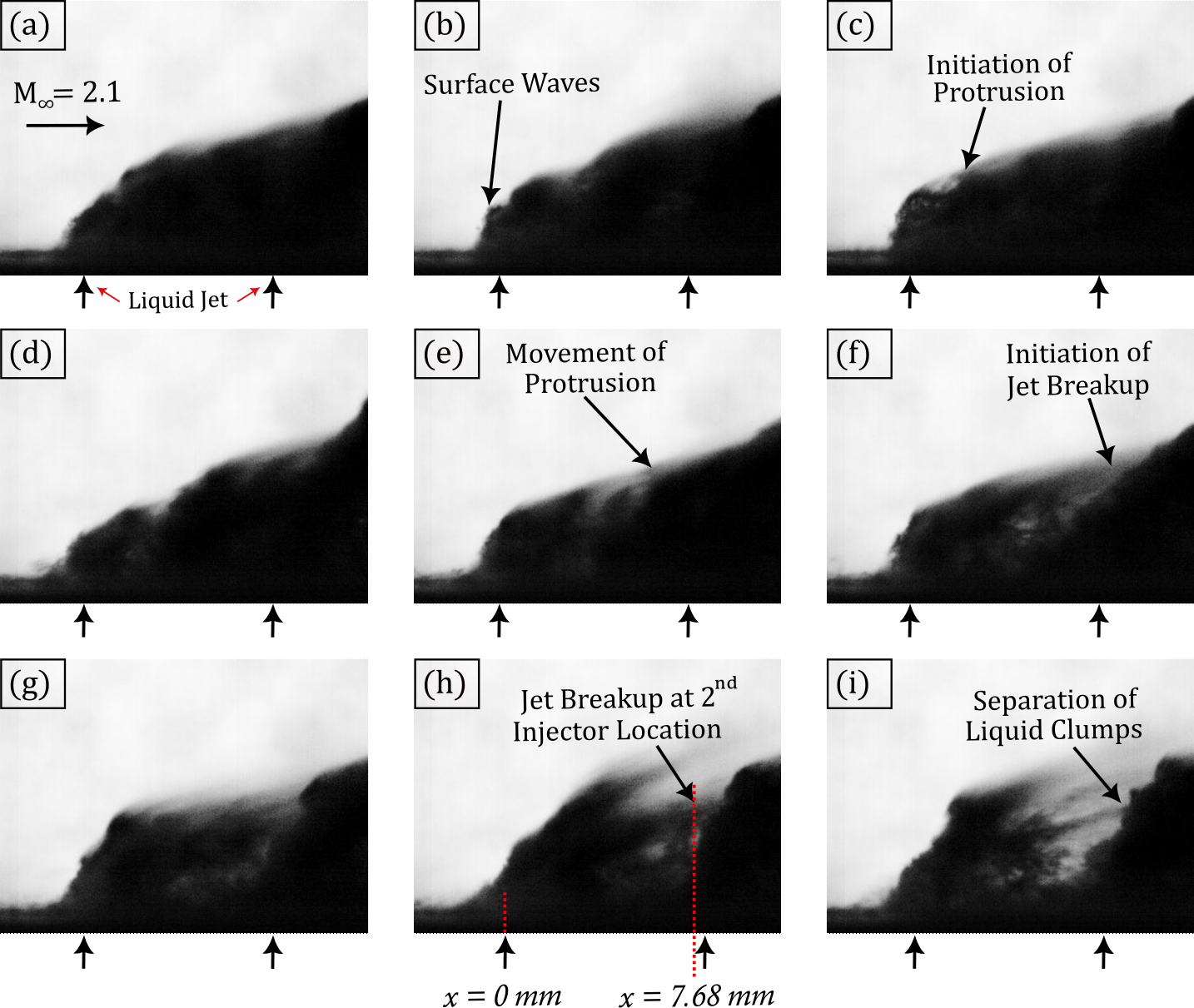}
  \caption{\label{fig:sequential_tndm}Sequential back-lit images captured with the tandem liquid injection at $J = 0.94$. (a) $t\textsubscript{0} = 0\ \mu s$ (b) $t\textsubscript{1} = 60\ \mu s$ (c) $t\textsubscript{2} = 120\ \mu s$ (d) $t\textsubscript{3} = 180\ \mu s$ (e) $t\textsubscript{4} = 240\ \mu s$ (f) $t\textsubscript{5} = 300\ \mu s$ (g) $t\textsubscript{6} = 360\ \mu s$ (h) $t\textsubscript{7} = 420\ \mu s$ (i) $t\textsubscript{8} = 480\ \mu s$}
\end{figure} 

The transient evolution of the injected jet with the tandem injection configuration, at the same \textit{J} value of 0.94, is shown in Fig. \ref{fig:sequential_tndm}. Similar to the findings observed in the single injection case, the tandem injection also demonstrates the formation of surface waves (shown in Fig. \ref{fig:sequential_tndm}(b)) and a characteristic protrusion structure (shown in Fig. \ref{fig:sequential_tndm}(c)) traveling downstream, eventually leading to the breakup of the liquid jet (shown in Fig. \ref{fig:sequential_tndm}(h)). A key difference observed between the tandem and single liquid injection is the location of the liquid jet breakup. For the tandem injection case, the breakup occurs just above the second injector location, approximately $x = 7.68\ mm$ downstream from the first injector. In contrast, for the single injection case, the liquid jet breakup is observed at a much greater distance, at around $x = 9.45\ mm$ from the injection location.

\begin{figure}[!htbp]
\includegraphics[width=0.47\textwidth]{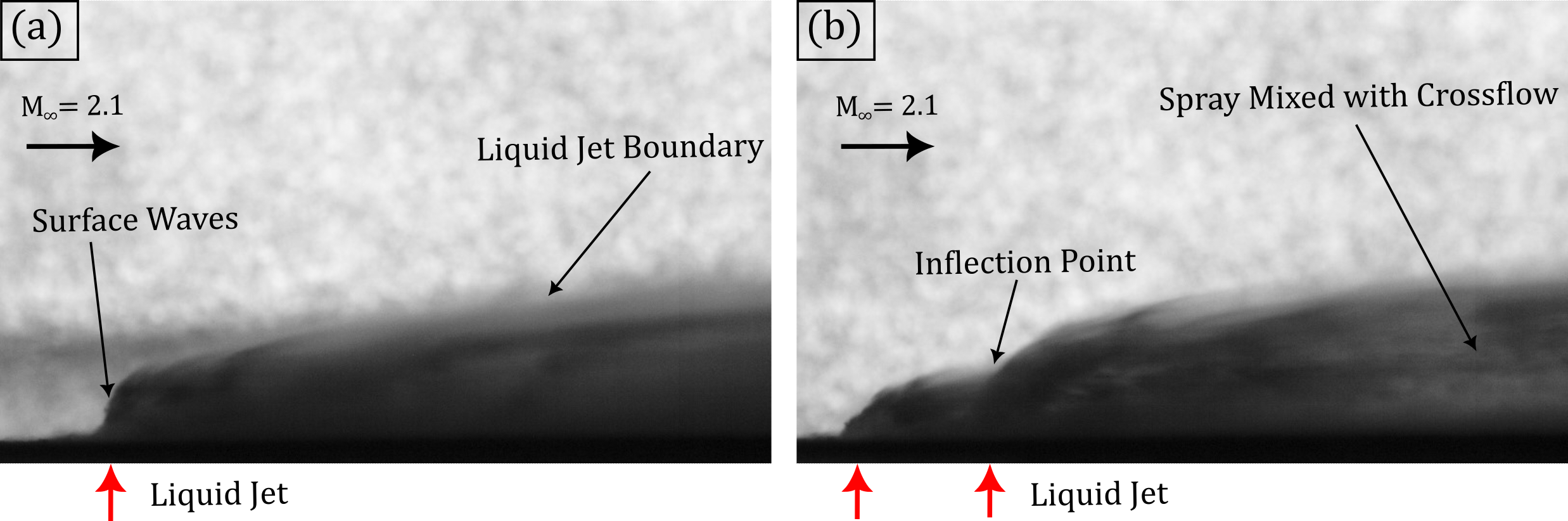}
  \caption{\label{fig:backlit_comp}Back-lit images captured at a particular time step for single and tandem injection configurations at $J = 0.94$. (a) Single Injection (b) Tandem Injection}
\end{figure}

Instantaneous back-lit images captured at a particular time instant with both single and tandem injection configurations have been shown and compared in Fig. \ref{fig:backlit_comp}(a) and Fig. \ref{fig:backlit_comp}(b), respectively. The liquid jet boundary and the surface waves can be clearly observed in Fig. \ref{fig:backlit_comp}(a). Additionally, it can be observed that the fine droplets mix effectively with the crossflow at a sufficient downstream distance from the injection location. A qualitative comparison between Fig. \ref{fig:backlit_comp}(a) and Fig. \ref{fig:backlit_comp}(b), reveals that there exists an inflection point at the jet boundary for the tandem injection case, located just above the second injection point. This inflection point signifies a change in the liquid jet trajectory due to the second injection in the tandem configuration, resulting in greater jet penetration compared to the single injection case.

\subsection{\label{sec:aero}Shock Wave Characteristics in Crossflow}
This section discusses the influence of pressurized liquid injection on the internal aerodynamics of a Mach 2.1 supersonic crossflow. Schlieren and back-lit imaging were employed to investigate the aerodynamic flow characteristics obtained with single as well as tandem injectors.

\begin{figure*}[htbp]
\includegraphics[width=0.92\textwidth]{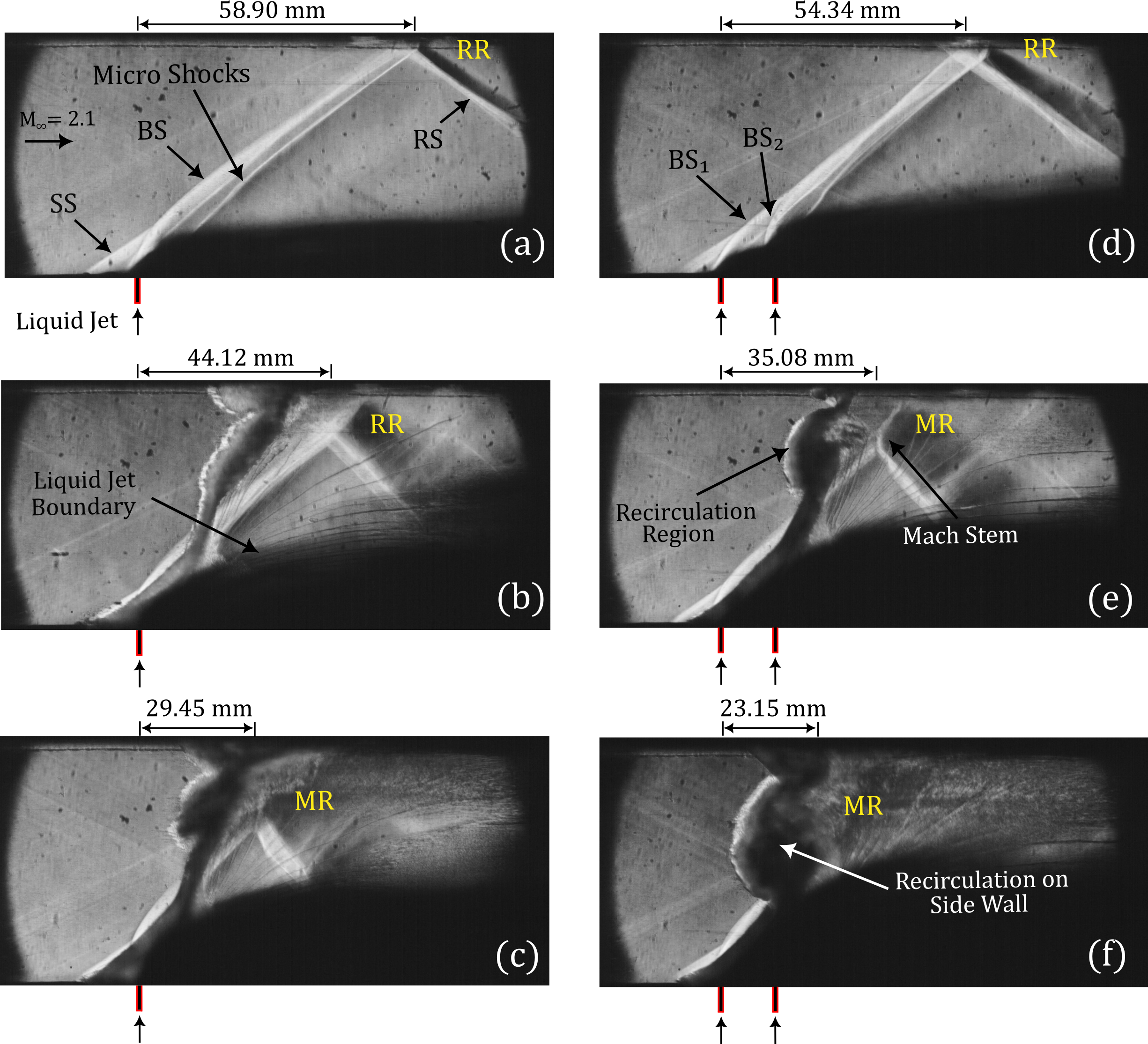}
  \caption{\label{fig:schlieren_comp}Schlieren images captured at a particular time step with the single injector (a, b, c) and tandem injector (d, e, f) under varying momentum flux ratios. (a) $J = 0.94$, Single (b) $J = 1.90$, Single (c) $J = 2.67$, Single (d) $J = 0.94$, Tandem (e) $J = 1.90$, Tandem (f) $J = 2.67$, Tandem}
\end{figure*} 

Fig. \ref{fig:schlieren_comp} shows the schlieren images of single and tandem injection configurations with various momentum flux ratios (\textit{J}) as labeled in Table \ref{tab:details_inj}. In Fig. \ref{fig:schlieren_comp}, the left side images (a, b, c) correspond to liquid injection through a single injector and the right side images (d, e, f) correspond to liquid injection through tandem ports with various momentum flux ratios of $J = 0.94$, $J = 1.90$, and $J = 2.67$, respectively. The general shock structures such as the incident bow shock wave (BS), micro shocks formed due to the surface waves of the liquid jet, separation shock formed upstream to the bow shock (SS), and the reflection of the bow shock wave from the top wall (RS) can be clearly seen in Fig. \ref{fig:schlieren_comp}(a). A comparison of Fig. \ref{fig:schlieren_comp}a and \ref{fig:schlieren_comp}(d) shows that the tandem injection case exhibits two bow shock waves in contrast to the single bow shock wave formed with the single injection. The secondary bow shock forms due to the presence of the second jet injection positioned downstream to the first jet. The bow shock waves generated due to the first and second injector are depicted as 'BS\textsubscript{1}' and 'BS\textsubscript{2}', respectively in Fig. \ref{fig:schlieren_comp}(d). This secondary bow shock wave (BS\textsubscript{2}) coalesces with the primary bow shock wave (BS\textsubscript{1}), forming a single bow shock that interacts with the top wall as seen in Fig. \ref{fig:schlieren_comp}(d).

Further comparison of Fig. \ref{fig:schlieren_comp}(a) and Fig. \ref{fig:schlieren_comp}(d) reveals that, at $J = 0.94$, the interaction of the bow shock wave with the top wall leads to a regular reflection for both the single and tandem injection. Moreover from Fig. \ref{fig:schlieren_comp}(d), it can be noted that, in the tandem injection case, the interaction of the bow shock wave with the top wall results in a larger incipient separation region/boundary layer thickening due to shock wave-boundary layer interaction compared to the single injection case with same momentum flux ratio. Fig. \ref{fig:schlieren_comp}(b) and \ref{fig:schlieren_comp}(e) show the schlieren images captured with single and tandem injection, respectively, at a higher momentum flux ratio of $J = 1.90$. It is seen that, despite the increase in the momentum flux ratio from $J = 0.94$ to $J = 1.90$, the single injection case still exhibits a regular reflection structure from the top wall as seen in Fig. \ref{fig:schlieren_comp}(b). It is also observed that, as the momentum flux ratio increases (from $J = 0.94$ to $J = 1.90$), the incipient separation region/boundary layer thickening near the interaction region at the top wall becomes more pronounced for $J = 1.90$ compared to the case with the momentum flux ratio of $J = 0.94$. However, with the same momentum flux ratio ($J = 1.90$), the bow shock wave reflection from the top wall produced by the tandem injection case shows a transition from regular reflection to Mach reflection as seen in Fig. \ref{fig:schlieren_comp}(e). From Fig. \ref{fig:schlieren_comp}(b) and Fig. \ref{fig:schlieren_comp}(e), it is observed that a patch of liquid stream travels from the bottom wall to the top along the side wall. For the tandem injection case, this liquid stream shows a large recirculation-type flow on the side wall, indicating a flow separation produced by the Mach stem of the MR as shown in Fig. \ref{fig:schlieren_comp}(e). A detailed discussion on the liquid flow along the sidewalls and the formation of the recirculation zone is presented in section \ref{sec:separation_aero}. With a further increase in momentum flux ratio to $J = 2.67$, bow shock wave reflection from the top wall exhibits a Mach reflection structure for both single and tandem injection as shown in Fig. \ref{fig:schlieren_comp}(c) and Fig. \ref{fig:schlieren_comp}(f), respectively. In Fig. \ref{fig:schlieren_comp}(f), the shock reflection patterns are obscured by the presence of a significant liquid patch and a recirculation region along the sidewall. However, the existence of a large recirculation zone on the side wall indicates the presence of a Mach reflection shock structure. In summary, the bow shock wave reflection from the top wall transitions from RR to MR at a higher momentum flux ratio for the single injection configuration compared to the tandem injection case. A detailed discussion on the mechanism of the shock transition from regular reflection to Mach reflection with an increase in the momentum flux ratio and injection configuration is given in the subsequent section \ref{sec:transition}.

\subsection{\label{sec:transition}Mechanism of Shock Transition from Regular Reflection to Mach Reflection}

The flow physics leading to the shock transition from RR to MR with higher momentum flux ratios has been investigated due to its immense implications in scramjet internal aerodynamics. It is well understood that the RR $\Leftrightarrow$ MR transition is primarily influenced by three factors: (1) the incoming flow Mach number, (2) incident shock wave angle, and (3) downstream pressure, and the details can be found in the monograph by \citet{ben2007shock}. In the current study, the C-D nozzle used has a constant area ratio (ratio of the nozzle exit area to the nozzle throat area) of 2 and the stagnation pressure inside the settling chamber is maintained the same for all the cases. Therefore the Mach number of the supersonic crossflow remains constant for all the experimental cases. The RR to MR transition in the present case, can hence be attributed to either an increase in the incident shock wave angle or an increase in the downstream pressure.\\

The effect of shock wave angle has been investigated by measuring the incident bow shock wave angle very close to the interaction location with the top wall for various cases using the Schlieren images (refer to Fig. \ref{fig:schlieren_comp}). MATLAB software has been used for the edge detection and the pixel data corresponding to the bow shock wave is extracted and the bow shock wave angle is computed. It is found that the incident bow shock wave angle very close to the interaction location of the bow shock wave with the top wall is around 32.8\textdegree\ and 37\textdegree\  for single and tandem injection configuration, respectively, at momentum flux ratios of $J = 0.94$ (Fig. \ref{fig:schlieren_comp}(a) and Fig. \ref{fig:schlieren_comp}(d)). For a moderate momentum flux ratio of $J = 1.90$, the bow shock wave angle measured for single and tandem injection configuration was found to be 36.8\textdegree\ and 42.3\textdegree\, respectively (Fig. \ref{fig:schlieren_comp}(b) and Fig. \ref{fig:schlieren_comp}(e)). However, for the case with a higher momentum flux ratio ($J = 2.67$), the shock wave angles could not be measured properly owing to the liquid traveling along the sidewall and obstructing the view of the shock structures for single and tandem injection as seen in Fig. \ref{fig:schlieren_comp}(c) and Fig. \ref{fig:schlieren_comp}(f), respectively. Nevertheless, the bow shock angle data for the momentum flux ratios of $J = 0.94$ and $J = 1.90$ reveal that the incident bow shock wave angle for the tandem injection is higher compared to the single injection. The increase in the bow shock angle can also be observed from the reduction in the distance between the location of the bow shock interaction with the top wall and the injection location as marked in Fig. \ref{fig:schlieren_comp}. The increased bow shock wave angle for the tandem injection compared to the single injection can be attributed to the higher penetration of the liquid jet which necessitates a higher deflection of the supersonic crossflow \citep{sathiyamoorthy2020penetration}. 
To investigate this further, the liquid jet penetration data was computed from the Schlieren images captured for various cases. The jet penetration data was determined by extracting the data points of the liquid jet boundary using MATLAB. To ensure accuracy and reduce the errors caused by inherent oscillations during liquid jet injection, a time averaged image was employed for data extraction. This averaged image was generated by averaging 100 consecutive images. The resulting averaged image is converted to a binary image and the canny edge algorithm is applied to accurately detect the liquid jet boundary. The extracted pixel data is converted into actual dimensions and subsequently normalized by the injector diameter to enable a consistent comparison for different cases. The liquid jet penetration has been compared for single and tandem injection configurations with two different momentum flux ratios. 

\begin{figure*}[htbp]
\includegraphics[width=1\textwidth]{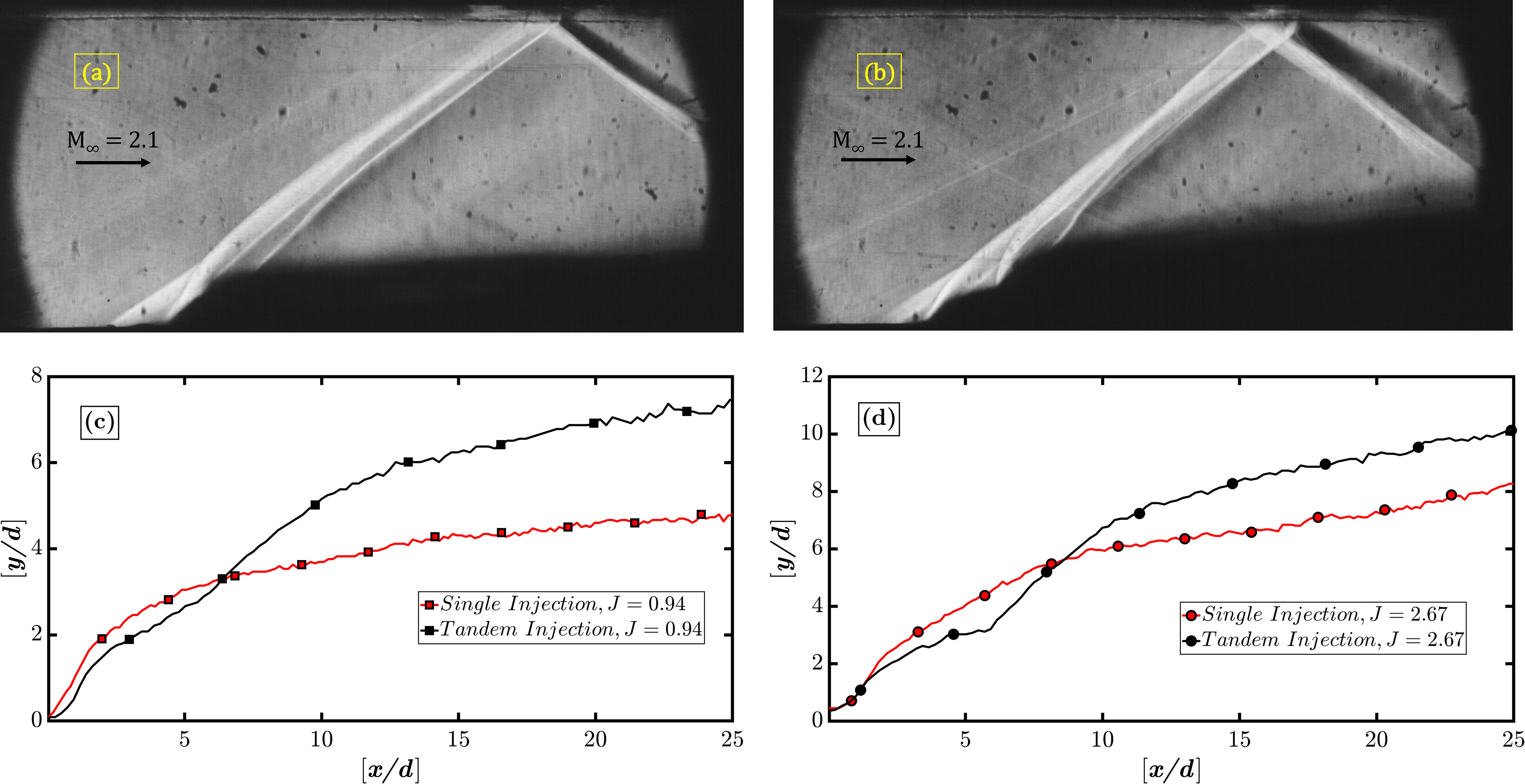}
\caption{\label{fig:single_tndm_penet}(a) Schlieren image with single injection at $J = 0.94$ (b) Schlieren image with tandem injection at $J = 0.94$ (c) Comparison of liquid jet penetration for single and tandem liquid injection at $J = 0.94$ (d) Comparison of liquid jet penetration for single and tandem liquid injection at $J = 2.67$}
\end{figure*}

Fig. \ref{fig:single_tndm_penet} provides both qualitative and quantitative comparison of the liquid jet boundary for the momentum flux ratios of $J = 0.94$ and $J = 2.67$. Fig. \ref{fig:single_tndm_penet}(a) and Fig. \ref{fig:single_tndm_penet}(b) present the instantaneous Schlieren images, providing a qualitative comparison of the liquid jet boundaries for single and tandem injection configurations, respectively for the momentum flux ratio of $J = 0.94$. It is evident from the Schlieren images that the liquid jet penetration is higher for the tandem injection compared to the single injection configuration with both having the same momentum flux ratio. The increased jet penetration for the tandem injection is due to the shielding effect provided by the first jet of the tandem injection. Fig. \ref{fig:single_tndm_penet}(c) and Fig. \ref{fig:single_tndm_penet}(d) show the normalized liquid jet boundary for both the injection configurations at $J = 0.94$ and $J = 2.67$, respectively which provides a quantitative understanding of the liquid jet penetration. It can be observed from Fig. \ref{fig:single_tndm_penet}(c) and \ref{fig:single_tndm_penet}(d) that the liquid jet penetration with the tandem injection is greater compared to the single injection configuration. A close observation of Fig. \ref{fig:single_tndm_penet}(c) and Fig. \ref{fig:single_tndm_penet}(d) shows that up to a certain critical distance from the injection location ($x/d = 6.4$ and $x/d = 8.6$ for $J = 0.94$ and $J = 2.67$, respectively), the single injection case shows a higher liquid jet penetration compared to the tandem injection case. After this critical distance, the tandem injection penetration shows a higher value compared to the single injection. The mass flow rate injected from each orifice in the tandem injection case is equal to half of the total injection mass flow rate for the single injection case since the total injected mass flow rate is the same for both single and tandem injection cases. Since only half of the total mass flow rate is introduced through the first injector in the tandem injection case, the jet penetration near the first injector is lesser due to a lower momentum flux ratio compared to the single injection case. However, in the tandem injection, the liquid jet from the second injector can penetrate a larger transverse distance into the crossflow due to the shielding effect produced by the first jet. The liquid injection from the first injector results in a bow shock wave (BS\textsubscript{1}) which reduces the crossflow momentum experienced by the second jet. This reduction in momentum results in a lesser aerodynamic force acting on the second jet, effectively increasing its penetration distance downstream. Additionally, the second jet impinges on the primary jet thereby contributing to an overall increase in the liquid jet penetration downstream. Therefore, an inflection point in the jet penetration trajectory can be observed for the tandem injection case compared to the single injection as shown in Fig. \ref{fig:single_tndm_penet}(c) and Fig. \ref{fig:single_tndm_penet}(d). In summary, the liquid jet penetration with the tandem injection is higher compared to the single injection for the same momentum flux ratio. This results in a higher bow shock wave angle for the tandem injection configuration for the same momentum flux ratio compared to a single injection which could be one of the possible reasons for the transition from RR to MR.

\begin{figure*}[htbp]
\includegraphics[width=1\textwidth]{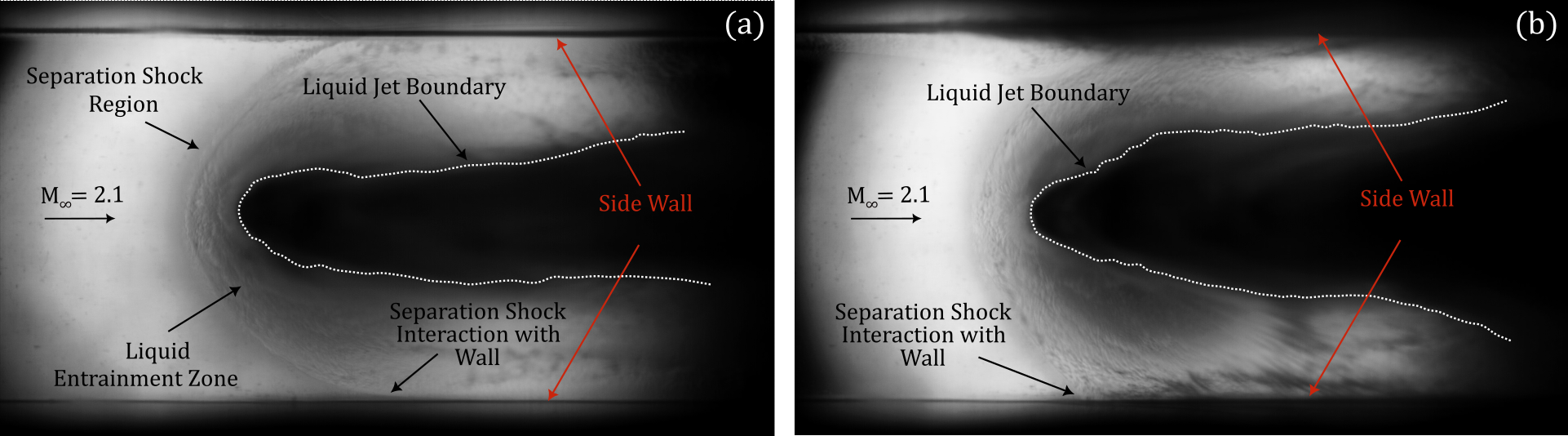}
\caption{\label{fig:top_view}Back-lit image captured from the top at a specific time instant using different injectors at a momentum flux ratio of $J = 2.67$ (a) Single Injector (b) Tandem Injector}
\end{figure*}

Another contributing factor for the RR to MR transition is the increase in the downstream pressure within the test section which can result from the reduction of the overall area for the passage of the supersonic crossflow. The area of the liquid jet at any location can be assessed by observing the penetration and the spanwise spread of the liquid jet. As previously observed, the liquid penetration is higher for the tandem injection configuration compared to a single injection for the same momentum flux ratio. Similarly, the spanwise spread is found to be higher with tandem injection configuration compared to the single injection for the same momentum flux ratio. This can be clearly seen from the back-lit imaging from the top side for the single and tandem injection case with the momentum flux ratio of $J = 2.67$ as shown in Fig. \ref{fig:top_view}(a) and Fig. \ref{fig:top_view}(b), respectively. The boundary of the liquid jet is clearly visible in Fig.\ref{fig:top_view}(a) and \ref{fig:top_view}b, outlined with a white dotted line. A qualitative comparison of Fig. \ref{fig:top_view}(a) and Fig. \ref{fig:top_view}(b) reveals that the spanwise spread of the liquid jet is greater for the tandem injection configuration compared to the single injection. A higher spanwise spread for the tandem injection case may result from the interaction between the two jets, where the second jet impinges on the first jet, causing the first jet to split and wrap around the second jet. It is also seen from Fig. \ref{fig:top_view}(a) and Fig. \ref{fig:top_view}(b), that the jet spread near the first injection location is smaller for the tandem injection case compared to the single injection case. This can be attributed to the lesser mass flow rate injected through the first orifice of the tandem injection case compared to the single injection as discussed previously. Further comparison between Fig. \ref{fig:top_view}(a) and Fig. \ref{fig:top_view}(b) indicates that, in the case of tandem liquid injection, the location at which the separation shock interacts with the side wall shifts upstream compared to the single injection case, owing to the larger deflection of the crossflow due to the larger spread of the jet in the tandem injection.

\begin{figure}[htbp]
\includegraphics[width=0.47\textwidth]{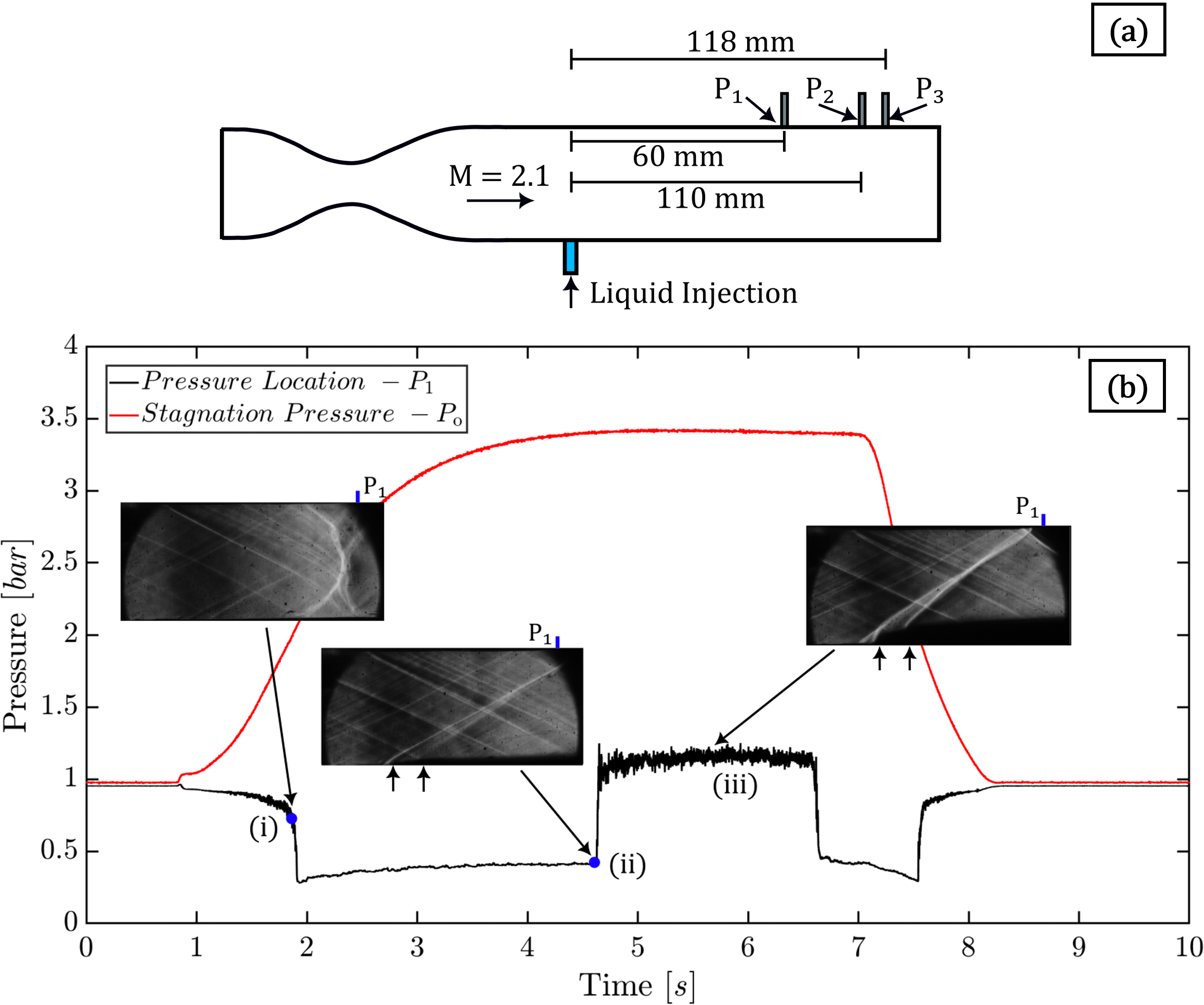}
\caption{\label{fig:sample_plot} (a) Schematic of the test section showing the pressure measurement locations (P\textsubscript{1}, P\textsubscript{2}, and P\textsubscript{3}) (b) Stagnation pressure (P\textsubscript{o}) measured within the settling chamber and the unsteady static pressure data measured at P\textsubscript{1} location for the tandem injection with a momentum flux ratio of $J = 0.94$}
\end{figure}

From the analysis of liquid jet penetration and spanwise spread, it is evident that both the penetration and spanwise spread are greater for the tandem injection configuration compared to the single injection configuration at the same momentum flux ratio. The combined increase in penetration and spanwise spread indicates that the cross-sectional area of the injected jet increases at any downstream location when the tandem injector is employed. Consequently, with the tandem injection, the effective passage area of the supersonic crossflow is reduced. In a supersonic flow, this reduction in the passage area acts as a diffuser, leading to a rise in the downstream static pressure. This increase in pressure can also lead to the transition of the bow shock wave reflection at the top wall from regular reflection to Mach reflection. To further investigate this, a detailed pressure measurement has been carried out in the downstream locations (P\textsubscript{1}, P\textsubscript{2}, and P\textsubscript{3} located at 60 \textit{mm}, 110 \textit{mm}, and 118 \textit{mm}, respectively, from the injection location) for various cases as shown in Fig. \ref{fig:sample_plot}(a).

Fig. \ref{fig:sample_plot}(b), shows the typical unsteady static pressure variation at the measurement location P\textsubscript{1} and the stagnation pressure (P\textsubscript{o}) variation in the settling chamber of the supersonic wind tunnel during an experimental cycle. In Fig. \ref{fig:sample_plot}(b) the red line indicates the stagnation pressure variation and the black line indicates the static pressure variation at P\textsubscript{1} location. The rise in the stagnation pressure indicates the opening of the valve of the wind tunnel and the stagnation pressure inside the settling chamber gradually rises and reaches a value of 3.4 \textit{bar}. This pressure remains nearly constant for a duration of 3.5 \textit{s} before the valve is closed. Simultaneously, the static pressure within the test section exhibits variations at different time instants, indicating distinct flow conditions during the operation. When the valve of the wind tunnel is opened, the static pressure at the measuring location P\textsubscript{1} shows a gradual decrement until $t = 1.89\ s$, after which a sudden dip in the static pressure can be observed (marked as '(i)'). The sudden dip in the static pressure is due to the passing of the starting normal shock at the measuring location P\textsubscript{1}. This can be clearly seen from the Schlieren image taken at the time instant corresponding to the sudden dip in the static pressure value as shown in the inset in Fig. \ref{fig:sample_plot}(b). After the passage of the starting shock, the static pressure shows a near-constant value of 0.38 \textit{bar} at the measurement location P\textsubscript{1} till $t = 4.63\ s$. At around $t = 4.63\ s$, a sudden rise in the pressure can be seen (marked as '(ii)') and this can be attributed to the initiation of the liquid jet injection. The injection of the liquid jet leads to the formation of a bow shock wave, causing an increase in the downstream pressure at the measuring location. The corresponding Schlieren image can be seen in the inset shown in Fig. \ref{fig:sample_plot}(b). Thereafter, the mean pressure remains nearly constant having a value of 1.1 \textit{bar} until $t = 6.59\ s$ (marked as '(iii)'). The pressure measurements in the region (iii) show an oscillatory nature which can be attributed to the fluctuations in the liquid jet boundary. The corresponding Schlieren image can be seen in the inset shown in Fig. \ref{fig:sample_plot}(b). A sudden decrease in the pressure can be observed at $t = 6.59\ s$ after region - iii, which marks the closing phase of the liquid injection. During the shutdown of the wind tunnel, a rapid rise in the static pressure is observed as the starting shock moves upstream through the test section. In the present experiments, the test section pressure corresponding to the region (iii) has been extracted from the unsteady pressure measurements and has been used for further investigations.

\begin{figure}[htbp]
\includegraphics[width=0.47\textwidth]{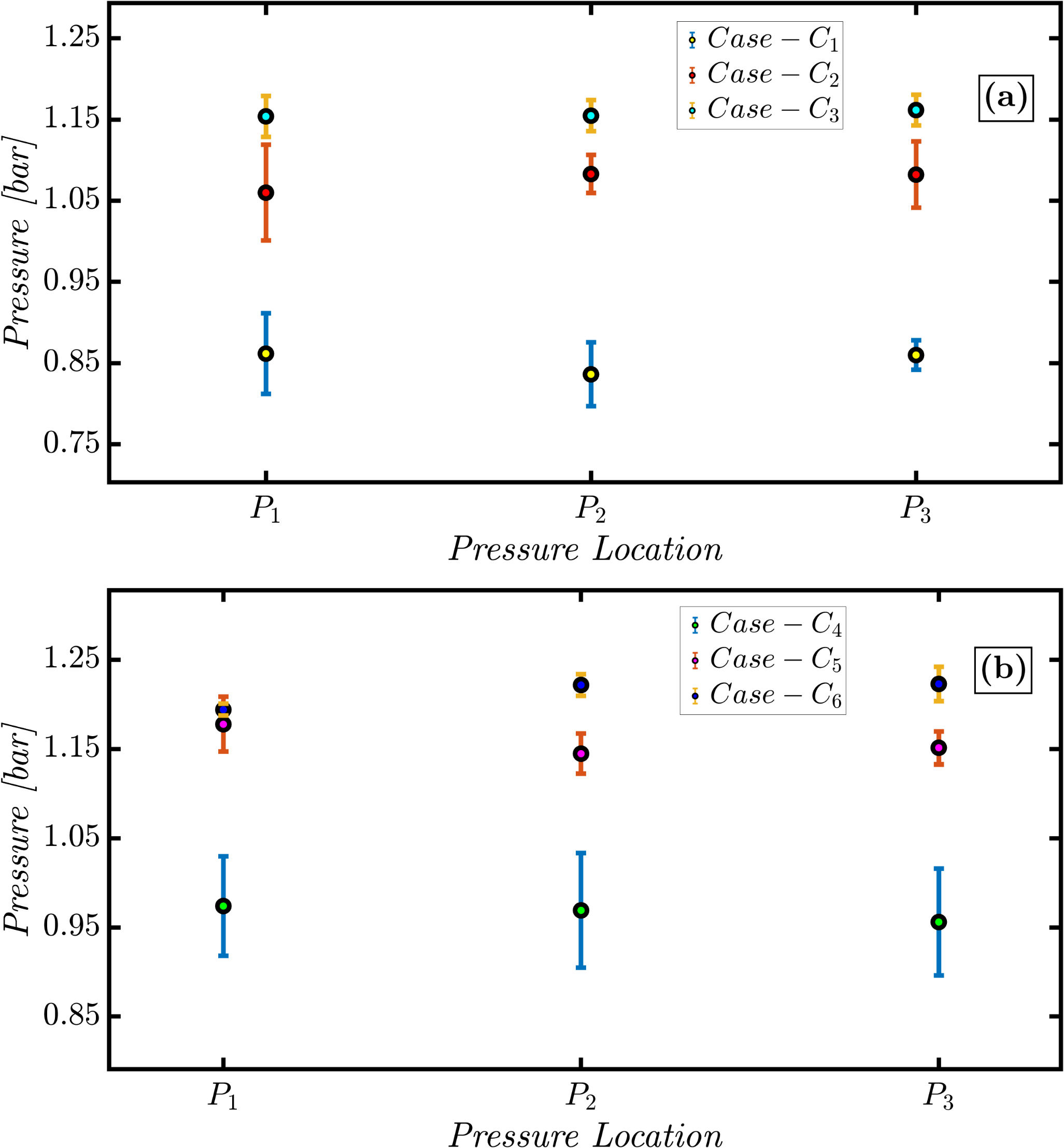}
\caption{\label{fig:single_tndm_pressure}Comparison of averaged static pressure measured at different locations P\textsubscript{1}, P\textsubscript{2}, and P\textsubscript{3} for both single and tandem injection configurations at different momentum flux ratios (a) Single Injection (C\textsubscript{1}, C\textsubscript{2}, and C\textsubscript{3}) (b) Tandem Injection (C\textsubscript{4}, C\textsubscript{5}, and C\textsubscript{6})}
\end{figure}

Since region (iii) shows large-scale pressure oscillations, the pressure measurement experiments have been repeated seven times for all the cases. 
The mean value at each location from each experimental run was calculated, followed by averaging the mean values across all runs for a specific case. These averaged results were then plotted to represent the data for specific momentum flux ratios and specific configurations. The static pressure data has been plotted and compared for three different momentum flux ratios with single and tandem injection configurations and is shown in Fig. \ref{fig:single_tndm_pressure}(a) and Fig. \ref{fig:single_tndm_pressure}(b), respectively.  Fig. \ref{fig:single_tndm_pressure}(a) illustrates that an increase in the momentum flux ratio results in a rise in the downstream static pressure at each measurement location. A similar trend is observed for the tandem injection configuration, as depicted in Fig. \ref{fig:single_tndm_pressure}(b).\\

\begin{figure}[htbp]
\includegraphics[width=0.47\textwidth]{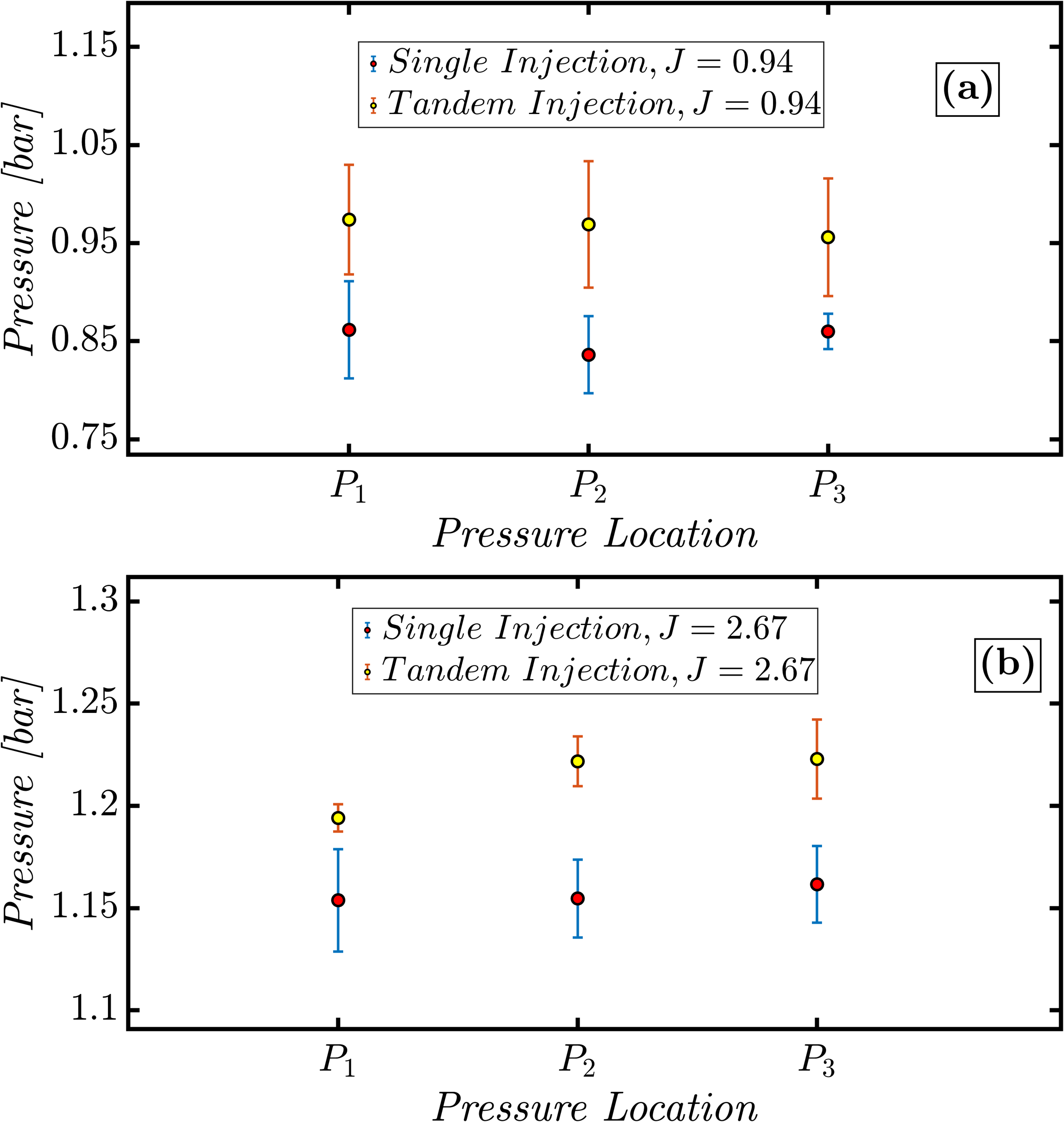}
\caption{\label{fig:mdot_single_tndm_pressure}Comparison of averaged static pressure measured at different locations for single and tandem injection configurations at different momentum flux ratios (a) $J = 0.94$ (b) $J = 2.67$}
\end{figure}

Further, the static pressure has also been compared for the single and tandem in a single plot for the momentum flux ratios of $J = 0.94$ and $J = 2.67$ which is shown in Fig. \ref{fig:mdot_single_tndm_pressure}(a), and Fig. \ref{fig:mdot_single_tndm_pressure}(b), respectively. As discussed earlier, the static pressure rises with the increase in the momentum flux ratio which can be clearly seen in Fig. \ref{fig:mdot_single_tndm_pressure}(a) and Fig. \ref{fig:mdot_single_tndm_pressure}(b). It can also be observed from the Fig. \ref{fig:mdot_single_tndm_pressure} that the static pressure at all the measurement locations (P\textsubscript{1}, P\textsubscript{2}, and P\textsubscript{3}) is higher for the tandem injection configuration compared to the single injection configuration at both the momentum flux ratios ($J = 0.94$ and $J = 2.67$). Therefore, the increased downstream static pressure within the test section for the tandem injection configuration compared to single injection at each of the momentum flux ratios can be attributed to an increase in both the liquid jet penetration and spanwise spread of the liquid jet. Based on the above observations, it can be concluded that the early transition from regular reflection to a Mach reflection in the tandem injection configuration compared to a single injection configuration, could be due to a combined effect of the increased incident bow shock wave angle at the interaction location with the top wall and an increase in the downstream static pressure within the test section.

\begin{figure*}[htbp]
\includegraphics[width=1\textwidth]{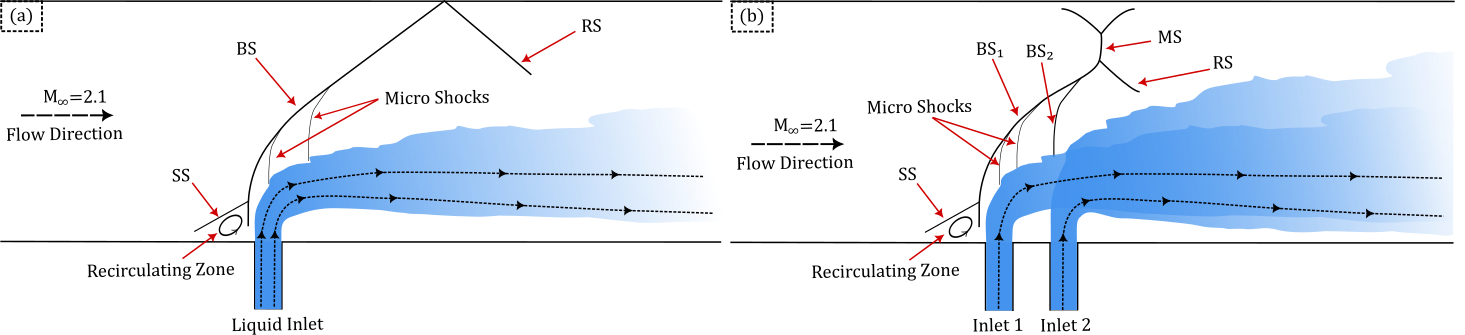}
\caption{\label{fig:ss_comp}Schematic of shock structures with different injectors (a) RR shock structure with Single Injector (b) MR shock structure with tandem Injector (SS - Separation Shock, BS - Bow Shock, RS - Reflected Shock, BS\textsubscript{1} - Bow Shock due to 1\textsuperscript{st} injector, BS\textsubscript{2} - Bow Shock due to 2\textsuperscript{nd} injector, MS - Mach Stem)}
\end{figure*}

\begin{figure*}[htbp]
\includegraphics[width=1\textwidth]{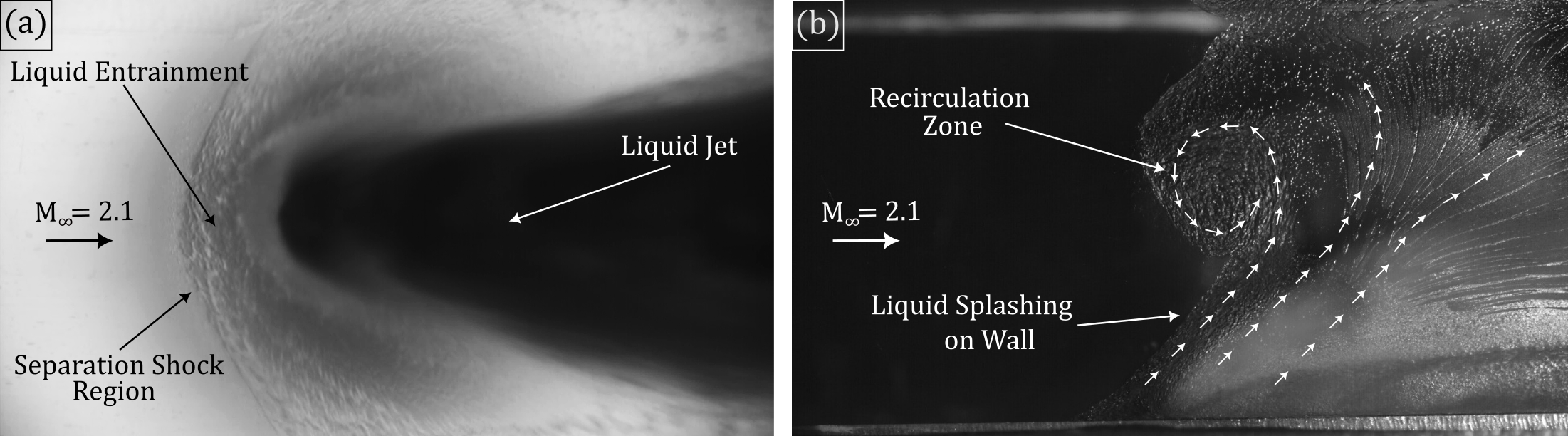}
\caption{\label{fig:side_view}Back-lit images captured from top view and side view at a specific time instant using tandem injector (a) Top View (b) Side View}
\end{figure*}

\subsection{\label{sec:separation_aero}Flow Separation Characteristics}
The schematic of shock wave interactions observed with single and tandem injectors at higher momentum flux ratios has been shown in Fig. \ref{fig:ss_comp}(a) and \ref{fig:ss_comp}(b), respectively. As discussed in section \ref{sec:aero}, at a momentum flux ratio of $J = 1.90$, the bow shock interaction with the top wall results in a regular reflection with the single injection. However, for the tandem injection with the same momentum flux ratio, the interaction of the bow shock wave with the top wall results in a transition from a regular reflection to a Mach reflection. It has to be noted that the Mach stem is a three-dimensional shock structure, and it extends and interacts with the side walls of the test section. The interaction of the Mach stem with the side wall creates an adverse pressure gradient across the Mach stem, thus, creating a local flow separation region and forming a recirculation zone on the side walls of the test section. Apart from the recirculation region formed on the side walls, a separation region is also formed on the bottom wall of the test section due to the bow shock wave interaction with the boundary layer. The separation regions on the bottom wall and the recirculation regions on the test section's side walls complicate the overall flow physics in such flows. This section discusses the observations and analysis of these separation regions.\\

\begin{figure*}[htbp]
\includegraphics[width=0.92\textwidth]{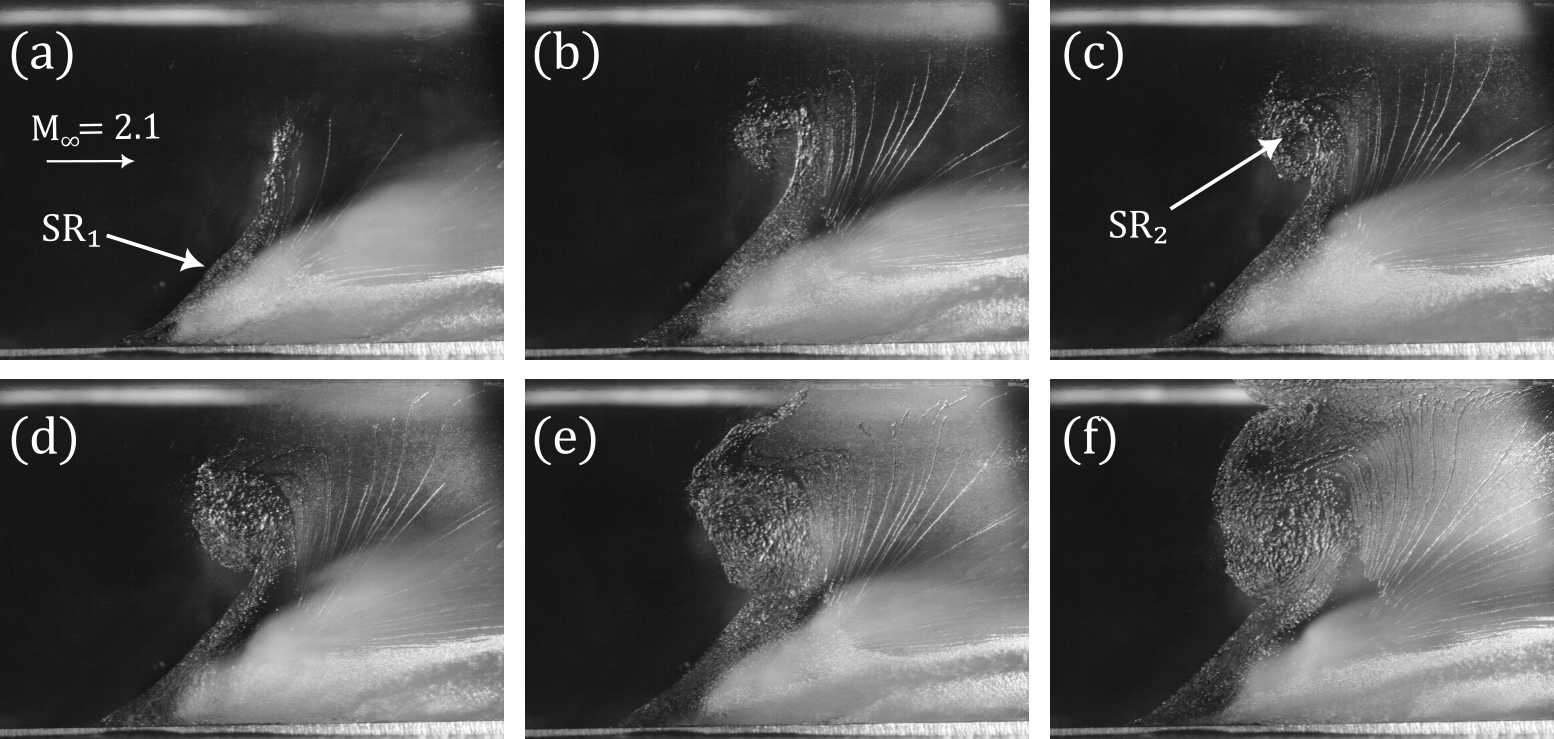}
\caption{\label{fig:sequential}Sequential back-lit images captured from side view using tandem injector at $J = 2.67$ (a) 145 \textit{ms} (b) 174 \textit{ms} (c) 180 \textit{ms} (d) 197 \textit{ms} (e) 215 \textit{ms} (f) 223 \textit{ms}}
\end{figure*}

To investigate the separation characteristics at the bottom wall due to the bow shock wave boundary layer interaction, a flow visualization study based on back-lit imaging has been carried out from the top side and is depicted in Fig. \ref{fig:side_view}(a). The flow visualization experiment was conducted with the tandem injection configuration at a momentum flux ratio of $J = 2.67$. As discussed in Section \ref{sec:Intro}, the interaction of the bow shock wave with the bottom wall results in flow separation due to the adverse pressure gradient created across the shock wave. This separation region produces an additional shock wave termed the 'separation shock' upstream of the separation bubble. This separation bubble is carried downstream due to the crossflow momentum, resulting in the formation of a horse-show vortex region upstream to the injection location \citep{li2019three}. The separation shock and the liquid jet boundary can be clearly seen in Fig. \ref{fig:side_view}(a). It is also observed that a fraction of the injected jet is getting entrained into the separation bubble due to the adverse pressure gradient across the shock wave. This region is termed the liquid entrainment region and is clearly shown in Fig \ref{fig:side_view}(a). A numerical study carried out by \citet{li2019three} reported that the liquid entrained in the separation region is carried downstream in the stream-wise direction by the horseshoe vortex formed in this region (refer Fig. \ref{fig:bowshock_schematic}). A similar phenomenon of liquid transport from the upstream region of the injection location is evident in Fig. \ref{fig:side_view}(a). The liquid entrained into the horseshoe vortex eventually interacts with the side walls and results in liquid splashing from the side walls of the test section. This liquid impinging on the side wall is carried upward along the bow shock wave interaction region with the side walls as shown in \ref{fig:side_view}(b). The back-lit imaging from the side (Fig. \ref{fig:side_view}(b)) also shows that the liquid stream traveling upward along the side wall is getting entrapped into a recirculation zone near the center of the side wall. The formation of this recirculation zone at the side walls of the test section can be attributed to the flow separation produced at the side wall due to the interaction of the Mach stem of the MR structure with the side wall. As reported previously, an increase in momentum flux ratio results in a Mach reflection when the bow shock wave interacts with the top wall. The Mach stem of this MR structure extends in the spanwise direction and interacts with the side wall. A strong adverse pressure gradient is produced near the Mach stem interaction region with the side wall. This leads to the formation of a local flow separation region at the side wall where the Mach stem interacts with the test section wall. This can also be confirmed from the Schlieren images shown in Fig. \ref{fig:schlieren_comp}(e), where the recirculation zone upstream to the Mach stem can be seen.

\begin{figure*}[htbp]
\includegraphics[width=0.9\textwidth]{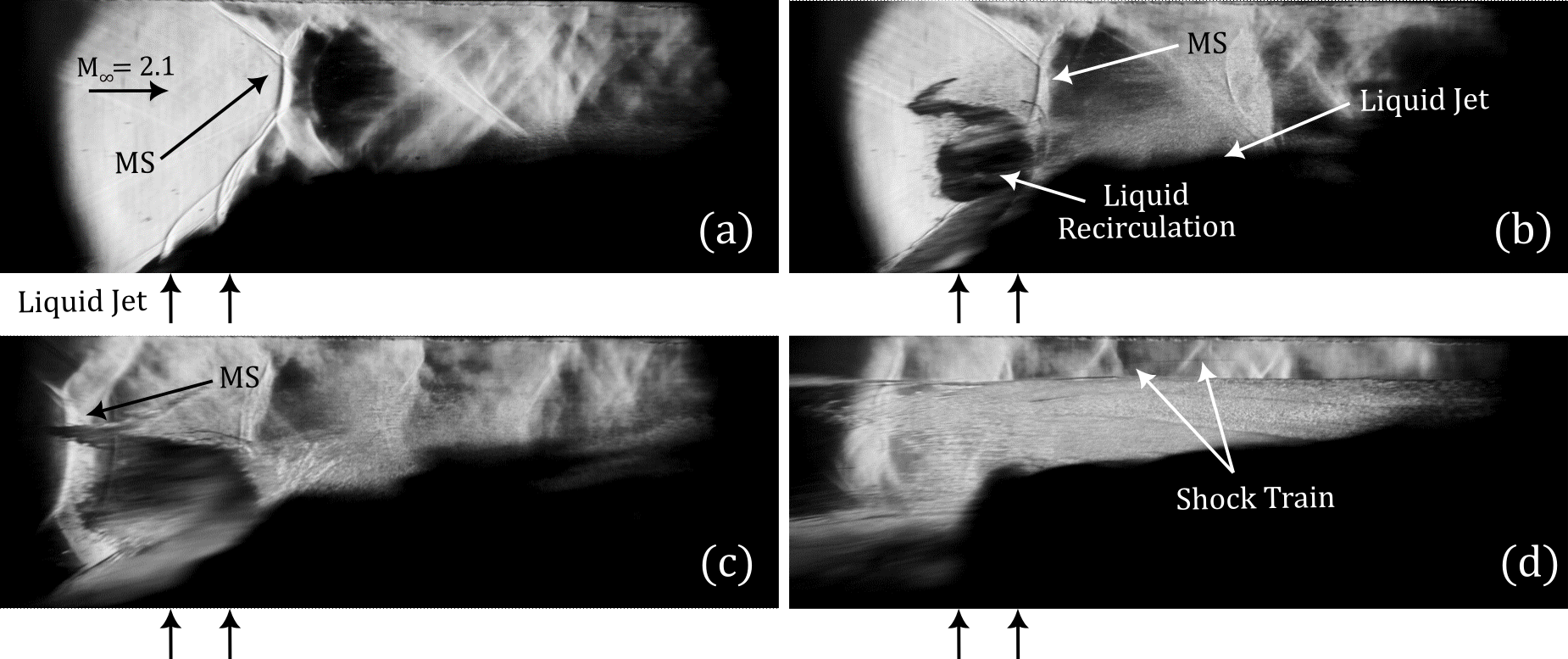}
\caption{\label{fig:unstart}Schlieren images captured for the tandem injection with a momentum flux ratio of $J = 2.81$ at various time instants (a) $t\textsubscript{1} = 136\ ms$ (b) $t\textsubscript{2} = 198\ ms$ (c) $t\textsubscript{3} = 230\ ms$ (d) $t\textsubscript{4} = 256\ ms$}
\end{figure*}

Fig. \ref{fig:sequential} shows the sequential back-lit images captured at different time instants for the tandem injection case with a momentum flux ratio of $J = 2.67$, revealing the transient evolution of the recirculation zone. Initially, a thin strip of separation region can be observed in the sidewall, denoted as 'SR\textsubscript{1}' in Fig. \ref{fig:sequential}(a). This thin separation region is formed due to the interaction of the incident bow shock wave with side walls, leading to shock boundary layer interaction and associated flow separation. The liquid, carried by the horseshoe vortex, interacts with the side wall and is entrained into this separation region (SR\textsubscript{1}) formed by the bow shock wave interaction with the side wall. The liquid then traverses upward through this separation region in the side wall eventually reaching the Mach stem location. At this point, the liquid encounters a large separation region (denoted as 'SR\textsubscript{2}' in Fig. \ref{fig:sequential}(c)) in the side wall produced by the Mach stem interaction with the wall. Subsequently, the liquid gets entrapped in this region, as depicted in Fig. \ref{fig:sequential}(b) through Fig. \ref{fig:sequential}(f). Additionally, it was observed that the liquid recirculation zone location shifted upstream, as the momentum flux ratio is increased.\\

An important feature observed with a momentum flux ratio of $J = 2.81$ is that the Mach reflection shock structure transiently moved upstream, leading to a possible tunnel unstart condition. This can be clearly seen in the sequential Schlieren images shown in Fig. \ref{fig:unstart}. As shown in Fig. \ref{fig:unstart}(a), captured at $t\textsubscript{1} = 136\ ms$ after the start of the injection, the Mach stem of the Mach reflection is positioned downstream of the injector location. Fig. \ref{fig:unstart}(b), shows the Schlieren image captured at $t\textsubscript{2} = 198\ ms$, where the Mach stem of the Mach reflection can be seen positioned directly above the injector location. The Schlieren images captured at $t\textsubscript{3} = 230\ ms$ and $t\textsubscript{4} = 256\ ms$ are shown in  Fig. \ref{fig:unstart}(c) and Fig. \ref{fig:unstart}(d), respectively, where it can be observed that the Mach reflection has traveled significantly upstream to the injector location, leaving only the shock trains visible in the captured view field. As discussed above, the Mach reflection structure produced by the liquid injection moves upstream and eventually stands at the exit of the C-D nozzle, when the liquid is injected at a higher momentum flux ratio. This is a potential case of scramjet engine unstart, leading to a sudden loss of thrust and possibly resulting in engine failure. Both single and tandem injection produced Mach reflection at a momentum flux ratio of $J = 2.67$, however, when the momentum flux ratio was increased further, both single and tandem injection scenarios encountered an unstart condition. The formation of the Mach reflection during the injection of the fuel causes a huge loss in the stagnation pressure due to the normal shock front of the Mach stem present in the Mach reflection shock structure.\\

The findings of this study highlight that the tandem injection enhances penetration and spanwise spread compared to the single injection. However, with the tandem injection configuration, the transition from regular reflection to Mach reflection was observed at a lesser momentum flux ratio compared to a single injection configuration. In summary, for the single injection configuration, liquid injection becomes disadvantageous in terms of the stagnation pressure loss at $J = 2.67$ or higher. In the case of tandem injection configuration, the liquid injection becomes disadvantageous in terms of the stagnation pressure at $J = 1.90$ or higher for the non-dimensional geometric parameters of the test section, $w/t = 1.6$, and $h/t = 2$, considered in the present study.

\section{\label{sec:conclusions}Conclusions}
The present study investigates the influence of injector configuration on the flow and aerodynamic characteristics during liquid injection in a supersonic crossflow within a confined duct. The experiments were conducted using two injector configurations: a single injector with a diameter of 2.83 \textit{mm} and a tandem injector, with each orifice having a diameter of 2 \textit{mm}. Three different injection mass flow rates were evaluated in this study. Schlieren visualization and high-speed imaging revealed that liquid column breakup begins with surface waves forming boundary protrusions, which fragment into large clumps and eventually finer droplets due to aerodynamic shear forces. The injection of the liquid into the supersonic crossflow results in the formation of a bow shock wave upstream of the injected jet. This bow shock wave interacts with the top wall resulting in shock reflections. At low momentum flux ratios (\textit{J = 0.90}), the reflection is regular (RR), while higher momentum flux ratios cause a transition to Mach reflection (MR). The tandem injection configuration transitions from RR to MR at a lower momentum flux ratio than the single injection case. This can be attributed to the greater penetration and spread of the injected jet in the tandem configuration. The greater penetration and spread reduce the effective cross-sectional area of the supersonic crossflow, causing a diffuser effect and an increase in downstream pressure. Additionally, the enhanced jet penetration requires a larger deflection of the crossflow, which increases the incident bow shock angle. These factors favor MR-type shock reflection, leading to an earlier transition in the tandem injection configuration compared to the single injection case. The bow shock wave's interaction with the bottom wall induces flow separation, entraining liquid into the separation zone. This entrained liquid is transported to the side wall by the horseshoe vortex generated in the separated flow. Upon interacting with the side wall, the bow shock wave drives the entrained liquid upward along the wall at the interaction point. In the case of an MR shock structure, the Mach stem reaches the side wall, causing flow separation due to the adverse pressure gradient. The entrained liquid is drawn into the resulting recirculation zone, creating a complex vortex structure near the side wall. Further investigations with single and tandem injection configurations with increased momentum flux ratios revealed that the Mach reflection structure gradually shifts upstream, severely compromising the performance and functionality of the scramjet engine.

\begin{acknowledgments}
The authors sincerely acknowledge the financial support provided by the Indian Space Research Organisation (ISRO), India, under Grant No. S/ISRO/AKR/20200104 for this project.
\end{acknowledgments}

\section*{Data Availability Statement}
The data that support the findings of this study are available from the corresponding author upon reasonable request.

\appendix*
\section{\label{sec:Mach_measure}Mach Number Measurement}

\begin{figure}[htbp]
\includegraphics[width=0.47\textwidth]{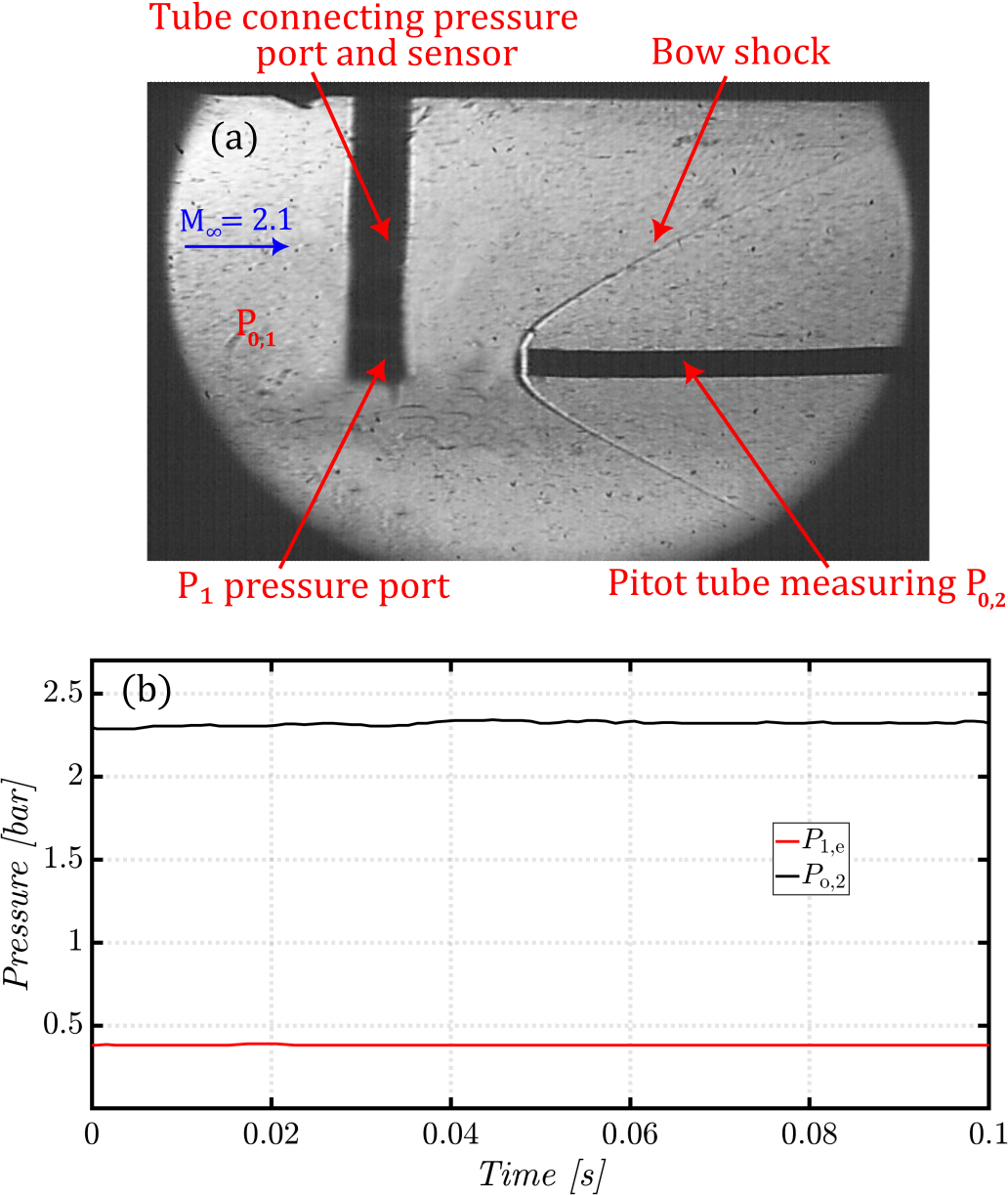}
\caption{\label{fig:Mach} (a) Schlieren image captured during the Mach number measurement (b) Variation of stagnation pressure downstream of the bow shock wave and static pressure upstream of the bow shock wave as a function of time }
\end{figure}

The convergent-divergent (C–D) nozzle was designed with an area ratio of 2, ($A/A\textsuperscript{*} = 2$) which corresponds to a theoretical Mach number of 2.2. The throat section and the exit section have a rectangular cross-section of $25\times 40\ mm\textsuperscript{2}$ (A\textsuperscript{*}) and $50\times 40\ mm\textsuperscript{2}$ (A), and the design of the C-D nozzle was carried out using the Method of Characteristics (MOC). The Mach number within the test section has been measured using the pitot stagnation tube and a static pressure port connected to a pressure sensor. A detached bow shock wave is formed upstream to the pitot tube, and the stagnation pressure downstream to the bow shock wave ($P\textsubscript{o,2}$) is measured by the pitot tube. The static pressure ($P\textsubscript{1,e}$) is measured using a circular port of 1 \textit{mm} diameter which is present upstream to the pitot tube and is connected to the pressure sensor using a tube as shown in Fig. \ref{fig:Mach}(a). The measured static pressure within the test section is found to be 0.38 bar. The variation of stagnation pressure downstream of the bow shock wave and static pressure upstream of the bow shock wave as a function of time is shown in Fig. \ref{fig:Mach}(b). The Mach number of the wind tunnel is calculated from the static-to-stagnation pressure ratio ($P\textsubscript{1}/P\textsubscript{o,2}$) using the normal shock relations \cite{zucker2019fundamentals}. To confirm flow uniformity, Mach number measurements were taken at multiple $y$-locations, and an average value was determined to represent the free-stream Mach number. The measurement uncertainty was assessed by repeating the experiments several times and calculating the standard deviation of the results. The average measured Mach number was found to be $2.1\pm 0.04$.

\section*{References}
%\nocite{*}
\bibliography{aipsamp}% Produces the bibliography via BibTeX.

\end{document}